\documentclass[iop]{emulateapj}
\slugcomment{{\it Recieved 2011 August 19; Accepted 2012 January 13}}
\usepackage{natbib,graphicx}
\begin{document}

\title{X-ray emission from an asymmetric blast wave and a massive white dwarf in the gamma ray emitting nova V407 Cyg}
\author{Thomas Nelson\altaffilmark{1,2,3}, Davide Donato\altaffilmark{1,4}, Koji Mukai\altaffilmark{1,2}, Jennifer Sokoloski\altaffilmark{5} and Laura Chomiuk\altaffilmark{6,7,8}}
\altaffiltext{1}{CRESST and X-ray Astrophysics Laboratory NASA/GSFC, Greenbelt, MD 20771, USA}
\altaffiltext{2}{Center for Space Science and Technology, University of Maryland Baltimore County, 1000 Hilltop Circle, Baltimore, MD 21250, USA}
\altaffiltext{3}{Current address: School of Physics and Astronomy, University of Minnesota, 115 Church St SE, Minneapolis, MN 55455}
\altaffiltext{4}{University of Maryland College Park, College Park, MD, USA}
\altaffiltext{5}{Columbia Astrophysics Laboratory, 550 W. 220th St, 1027 Pupin Hall, Columbia University , New York, NY 10027, USA}
\altaffiltext{6}{Harvard-Smithsonian Center for Astrophysics, 60 Garden St, Cambridge, MA 02138, USA}
\altaffiltext{7}{National Radio Astronomy Observatory, P.O. Box 0, Socorro, NM 87801, USA}
\altaffiltext{8}{Jansky Fellow}
\email{thomas.nelson@nasa.gov}

\begin{abstract}
Classical nova events in symbiotic stars, although rare, offer a unique opportunity to probe the interaction between ejecta and a dense environment in stellar explosions.  In this work, we use X-ray data obtained with {\it Swift} and {\it Suzaku} during the recent classical nova outburst in V407 Cyg to explore such an interaction.  We find evidence of both equilibrium and non-equilibrium ionization plasmas at the time of peak X-ray brightness, indicating a strong asymmetry in the density of the emitting region.  Comparing a simple model to the data, we find that the X-ray evolution is broadly consistent with nova ejecta driving a forward shock into the dense wind of the Mira companion.   We detect a highly absorbed soft X-ray component in the spectrum during the first 50 days of the outburst that is consistent with supersoft emission from the nuclear burning white dwarf.  The high temperature and short turn off time of this emission component, in addition to the observed breaks in the optical and UV lightcurves, indicate that the white dwarf in the binary is extremely massive.   Finally, we explore the connections between the X-ray and GeV $\gamma$-ray evolution, and propose that the gamma ray turn-off is due to the stalling of the forward shock as the ejecta reach the red giant surface.
\end{abstract}


\section{Introduction}
\label{intro}
Classical novae (CNe) are a subclass of cataclysmic variable star (white dwarfs accreting material from a binary companion) that experience rare, sudden increases in brightness of up to 9 magnitudes.  These outbursts are now understood to be the result of a thermonuclear runaway event in the shell of accreted material on the white dwarf surface, initiated once the temperature and pressure required for nuclear burning are reached.  A small number of nova systems have been detected in outburst more than once, and are referred to as recurrent novae.  However, all novae are expected to recur on timescales ranging from a less than a decade to 100,000s of years, depending primarily on the mass of the white dwarf ($M_{WD}$) and the mass accretion rate ($\dot{M}$).  More massive white dwarfs have higher surface gravities, and therefore higher pressures in the accreted shell - as a result, less mass is needed to reach the critical pressure required for the onset of the thermonuclear runaway.  Higher mass accretion rates decrease the time it takes to reach the critical accreted mass.  

Most classical nova events occur in close binary systems, where the mass donor is a Roche Lobe filling, late type main sequence star.  More rarely, the system is a wide binary hosting a white dwarf and a late type giant star, which may or may not fill its Roche Lobe.  Four of the known recurrent novae (RS Oph, T CrB, V745 Sco and V3890 Sgr) have giant companions.  These systems are also classified as symbiotic stars.  The nature of the mass donor has major implications for the evolution of the outburst over time, particularly in terms of the interaction of the nova ejecta and the circumbinary environment.

Over the last three decades, CNe in outburst have been established as an important class of X-ray emitting object.  They are perhaps best known as supersoft X-ray sources (SSS), characterized by blackbody like emission with 40 $<$ $kT~ (\rm{ eV})$ $<$ 80 and 10$^{36}$ $<$ $L_{X}~ ({\rm erg~s^{-1}})$ $<$ 10$^{38}$.  The SSS phase of the nova outburst occurs when the accreted envelope has expanded enough to become optically thin to soft X-rays, and thus we see the radiation originating in the burning layer.  CNe are also known to emit hard X-rays at early times, typically at much lower luminosities than associated with the SSS phase \citep{Mukai01,Mukai08}.  This emission has two possible origins.  In most CNe, where there is very little material surrounding the binary, the X-ray emission is thought to originate in internal shocks within the nova ejecta. More rarely, the X-rays are due to interaction of the nova ejecta with a dense circumbinary medium, as in RS Oph \citep{Sokoloski06,Bode06}. Events of this type can be considered analogous to supernovae in dense circumstellar media, whose early X-ray emission is also due to the interaction of the ejecta and the local environment \citep{Chevalier82,Fransson96}.  

Novae in symbiotic star systems offer a unique opportunity to study shock physics relevant for supernova remnant evolution.  The local nature of novae make them easier to study in X-rays, and the much smaller ejected masses involved allow the evolving interaction with the surrounding medium  to be observed over modest time periods. Symbiotic stars may also be progenitors of some type Ia supernovae.  In recent years, variable Na D absorption features have been detected in a small number of systems that have been interpreted as the presence of material around the progenitor system.\citep[see e.g.][]{Patat07,Simon09}.  \citet{Patat11} presented similar evidence for evolving Na D absorption in the 2006 outburst of RS Oph, and discussed the role of symbiotic novae as the progenitors of at least a subset of SNe Ia.

\subsection{The 2010 outburst of V407 Cyg}
V407 Cyg is a member of the small group of symbiotic Mira stars (binaries comprised of a white dwarf and a Mira-type AGB companion).  The system is only weakly symbiotic, with no detectable continuum from the hot component in quiescence in the optical or near UV bands, and only weak Balmer series emission lines.  The He II 4686 emission line is occasionally detected. The pulsation period of the Mira is 745 $\pm$ 15 days \citep{Munari90,Kolotilov98}.  This is the longest pulsation period of all known symbiotic Mira systems.  Lithium has been identified in the optical spectrum of V407 Cyg \citep{Tatarnikova03a}, indicating that this system may be experiencing so-called ``hot bottom burning" (HBB.)  This makes determining the distance to the system difficult; Mira variables with HBB tend to me more luminous than their period would suggest, so the standard period luminosity relation over estimates the distance.  Current estimates in the literature range from 1.9 kpc \citep{Kolotilov98} to 2.7 kpc \citep{Munari90}.  Throughout this work, we will adopt the Munari et al. distance.

The orbital parameters of the system are not constrained, although \citet{Munari90} claim an orbital period of 43 years based on possible dust obscuration events in the historic lightcurve.  If we assume that the white dwarf mass is $\sim$1 M$_{\odot}$, and that the Mira mass is in the range 4--8 M$_{\odot}$ since it is experiencing HBB \citep{Mikolajewska10}, then the 43 year period implies a separation of 20--25 AU.  

V407 Cyg has experienced two small amplitude outbursts (peak visual brightness $\sim$13 mag) in the past, in 1936 and 1998.  The 1936 event was originally characterized as a classical nova event \citep{Hoffmeister49}, although this interpretation seems unlikely given the most recent event.  The 1998 event has been referred to as a symbiotic type outburst.   The most recent outburst was discovered simultaneously by two Japanese groups on 2010 March 10 \citep{Hirosawa10,Nakano10}, with a maximum magnitude in unfiltered visual CCD images of 6.8 mags.  The amplitude of this outburst was much larger than the events observed in 1936 and 1998.    \citet{Munari11} presented optical and near infrared photometry obtained over 200 days following the outburst.  From those data, the authors obtained values of $t_{2}$ and $t_{3}$ (the time to fade from peak V band brightness by 2 and 3 magnitudes) of 5.9 and 24 days, respectively, making this outburst of V407 Cyg a fast nova event.  

\citet{Munari11} and \citet{Shore11} also presented optical spectroscopy from the first few months of the 2010 outburst.  The earliest spectra are characterized by bright emission lines of the Balmer series of hydrogen, and of He I.  The evolution of these line features was almost identical to that observed during the 2006 outburst of RS Oph.  After day 48, a system of coronal lines appeared typical of classical novae in outburst and again similar to the 2006 outburst of RS Oph. The high resolution spectra of \citet{Shore11} revealed asymmetric structure in several emission lines, with red wings extending to larger velocities than blue wings.  The authors attribute these features to the asymmetric density structure of the binary environment encountered by the nova ejecta.  The equivalent width of some line features (e.g. O I 8448) appears to evolve in tandem with the X-ray emission, while others (e.g. O I 6300) more closely follow the optical and UV evolution.   

The most remarkable aspect of the 2010 outburst of V407 Cyg was the detection of the nova as a transient $\gamma$-ray source with the LAT detector onboard the \textit{Fermi} observatory \citep{Abdo10}.   The first significant detection of the source was on 2010 March 10, the same day as the discovery of the optical outburst.  The \textit{Fermi} source is coincident with the optical position of V407 Cyg to within 0.040$^{\circ}$, well within the 95\% confidence positional error of the LAT.   Both the temporal and spatial coincidence with the V407 Cyg optical outburst indicate that this event is the origin of the transient $\gamma$-rays.  The peak $\gamma$-ray flux was observed on days 3 and 4 after the outburst, after which the source began to fade.  The last significant ($>$3$\sigma$) \textit{Fermi} detection was on day 15 of the outburst (2010 March 25).  Gamma-ray emission from novae has been predicted at energies $<$1 MeV due to the decay of radioactive isotopes of sodium and beryllium \citep{Hernanz05}, although this has not observed to date.  The {\it Fermi} detection of V407 Cyg has demonstrated that in some cases novae are also capable of producing $\sim$GeV photons, adding a new class of transient to the $\gamma$-ray sky.

In this article, we use X-ray observations obtained with {\it Swift} and {\it Suzaku} to study the interaction of the nova ejecta with the wind of the Mira component in V407 Cyg.  In Section 2, we discuss the available data and our reduction procedure.  Next, we present a brief optical, UV and X-ray overview of the outburst in Section 3 and a detailed spectral analysis of the \textit{Suzaku} data in Section 4.  Then, in Section 5, we analyze the \textit{Swift} XRT spectral evolution utilizing the results of our \textit{Suzaku} analysis.  In Section 6 we present a simple model of the outburst and utilize it to explore the gross features of the X-ray  lightcurve.  We compare the X-ray evolution of V407 Cyg with that of RS Oph in Section 7.  Finally, we discuss our results in Section 8, and present a summary and our main conclusions in Section 9.

\section{Observations and Data Reduction}
\label{data}

\subsection{\textit{Swift}}
\textit{Swift} began observing V407 Cyg as part of a target of opportunity (ToO) program just three days after the outburst.  The ToO program was then extended a number of times as the system continued to evolve.  The final dataset consists of 34 individual pointings spanning the first 120 days of the outburst.  In all but four observations, a UV image in the UVM2 filter was also obtained with the UVOT. The details of each observation, including exposure time, XRT count rates and UVOT magnitudes, are presented in Table 1.  An analysis of this dataset was also presented by \citet{Shore11}.

We performed the reduction and analysis of the \textit{Swift} data using a processing script customized for the XRT and UVOT data (Donato et al., in preparation).  The script re-processes the data, stored in the HEASARC archive, using the standard HEASoft software (version 6.8) and the latest calibration database (20091130 for XRT and 20100129 for UVOT).  The reduction of XRT data consists of running {\tt xrtpipeline}, selecting only events with grades 0--12 in photon counting mode (PC), while the UVOT event mode data are processed following the steps reported in the UVOT Software Guide 2.2.

The XRT data analysis is performed using the script {\tt xrtgrblc}, available in the HEASoft software package. In brief, the script selects the best source and background extraction regions based on the source intensity (typically a circle and an annulus, respectively). In the case of V407 Cyg, the source extraction region has a radius of up to 55\arcsec, and the background annulus has inner and outer radii of up to 80\arcsec\ and 135\arcsec, respectively, when the light curve is close to its peak count rate. If field sources are present in the background region, they are excluded using circular masks whose radius depends on the field source intensity.  Using these regions, source and background count rates, spectra and event lists are extracted. The net source count rate has to be corrected for irregularities in the exposure map and the Point Spread Function (PSF).  The total correction factor is obtained using the HEASoft tool {\tt xrtlccorr}.

The script that handles the UVOT analysis ({\tt uvotgrblc}) determines the presence of field sources and excludes them from the background region with circular regions whose size again depends on intensity. The optimal background region is chosen after comparing 3 annular regions centered on the main source in summed images. The region with the lowest background is selected. In the case of V407 Cyg, the field of view is contaminated by many CCD artifacts  (e.g., ghost images and/or readout streaks) from a nearby very bright star. For this reason, a custom background region has been chosen to minimize the 
effects of these artifacts, consisting of two 15\arcsec\ circular regions located 40\arcsec\ from V407 Cyg, one at 146\degr\ and the other at 326\degr\ position angle. 

The source extraction region is also intensity dependent: {\tt uvotgrblc} selects the aperture size between 3\arcsec\ and 5\arcsec\ based on the observed count rate and the presence of close field sources. V407 Cyg is a bright source even in the UVM2 filter ($\lambda_{cen}$ = 2246 \AA) and only dim field sources are present.  A 5\arcsec\ region is selected for every pointing. The script estimates the best source position using the task {\tt uvotcentroid} and the photometry using the task {\tt uvotsource}. The resulting values are corrected for aperture effects but not for interstellar extinction.  Magnitudes are presented in the AB system.

\begin{deluxetable*} {lrrcccc}[h]
\tabletypesize{\tiny}
\tablecaption{Summary of observations} 
\tablewidth{0pt}
\tablehead{
\colhead{Date} & \colhead{Days from $t_{0}$}\tablenotemark{a} & \colhead{$t_{exp}$ (s)} & \colhead{rate (0.3--10 keV) (c s$^{-1}$)} & \colhead{rate (0.3--2 keV) (c s$^{-1}$)}  & \colhead{rate (2--10 keV) (c s$^{-1}$)}  & \colhead{UVM2 mag}      
}
\startdata   
\cutinhead{\textit{Swift}}
2010-03-13 & 3.8 & 949  & 0.011 $\pm$ 0.004 & $<$0.012          & 0.008 $\pm$ 0.003 & ...             \\ 
2010-03-15 & 5.8  & 955  & 0.011 $\pm$ 0.003 & $<$0.008          & 0.010 $\pm$ 0.003 & 12.57 $\pm$ 0.03\\ 
2010-03-19 & 9.3  & 922  & 0.023 $\pm$ 0.005 & 0.008 $\pm$ 0.003 & 0.015 $\pm$ 0.004 & 12.71 $\pm$ 0.03\\ 
2010-03-23 & 13.2 & 956  & 0.009 $\pm$ 0.003 & $<$0.014          & $<$0.009          & 12.76 $\pm$ 0.03\\ 
2010-03-24 & 14.7 & 2169 & 0.016 $\pm$ 0.003 & 0.004 $\pm$ 0.001 & 0.013 $\pm$ 0.002 & ...             \\ 
2010-03-25 & 16.0 & 916  & 0.023 $\pm$ 0.005 & 0.007 $\pm$ 0.003 & 0.016 $\pm$ 0.004 & ...             \\ 
2010-03-27 & 17.2 & 976  & 0.048 $\pm$ 0.007 & 0.017 $\pm$ 0.004 & 0.031 $\pm$ 0.006 & 12.78 $\pm$ 0.03\\ 
2010-03-31 & 21.2 & 1237 & 0.158 $\pm$ 0.011 & 0.046 $\pm$ 0.006 & 0.113 $\pm$ 0.010 & 12.82 $\pm$ 0.03\\ 
2010-04-02 & 23.8 & 3164 & 0.210 $\pm$ 0.008 & 0.079 $\pm$ 0.005 & 0.130 $\pm$ 0.006 & 12.90 $\pm$ 0.02\\ 
2010-04-04 & 25.3 & 4194 & 0.209 $\pm$ 0.007 & 0.089 $\pm$ 0.005 & 0.121 $\pm$ 0.005 & 12.94 $\pm$ 0.02\\ 
2010-04-06 & 27.7 & 4217 & 0.223 $\pm$ 0.007 & 0.108 $\pm$ 0.005 & 0.114 $\pm$ 0.005 & 13.04 $\pm$ 0.02\\ 
2010-04-08 & 29.1 & 3981 & 0.244 $\pm$ 0.008 & 0.125 $\pm$ 0.006 & 0.120 $\pm$ 0.005 & 13.03 $\pm$ 0.02\\ 
2010-04-10 & 31.2 & 3871 & 0.263 $\pm$ 0.008 & 0.161 $\pm$ 0.006 & 0.104 $\pm$ 0.005 & 13.06 $\pm$ 0.02\\ 
2010-04-12 & 33.1 & 3002 & 0.244 $\pm$ 0.009 & 0.146 $\pm$ 0.007 & 0.097 $\pm$ 0.006 & 13.21 $\pm$ 0.02\\ 
2010-04-14 & 35.1 & 2003 & 0.235 $\pm$ 0.011 & 0.136 $\pm$ 0.008 & 0.108 $\pm$ 0.007 & 13.21 $\pm$ 0.02\\ 
2010-04-16 & 37.5 & 4801 & 0.225 $\pm$ 0.007 & 0.150 $\pm$ 0.006 & 0.075 $\pm$ 0.004 & 13.22 $\pm$ 0.02\\ 
2010-04-17 & 38.1 & 9828 & 0.219 $\pm$ 0.005 & 0.157 $\pm$ 0.004 & 0.063 $\pm$ 0.003 & ...             \\ 
2010-04-18 & 39.4 & 4740 & 0.218 $\pm$ 0.007 & 0.158 $\pm$ 0.006 & 0.062 $\pm$ 0.004 & 13.26 $\pm$ 0.02\\ 
2010-04-20 & 41.0 & 3357 & 0.211 $\pm$ 0.008 & 0.148 $\pm$ 0.007 & 0.062 $\pm$ 0.004 & 13.30 $\pm$ 0.02\\ 
2010-04-22 & 43.8 & 3602 & 0.173 $\pm$ 0.007 & 0.126 $\pm$ 0.006 & 0.046 $\pm$ 0.004 & 13.50 $\pm$ 0.02\\ 
2010-04-24 & 45.6 & 2784 & 0.149 $\pm$ 0.007 & 0.108 $\pm$ 0.006 & 0.041 $\pm$ 0.004 & 13.52 $\pm$ 0.02\\ 
2010-04-28 & 49.3 & 2725 & 0.128 $\pm$ 0.007 & 0.095 $\pm$ 0.006 & 0.034 $\pm$ 0.004 & 13.86 $\pm$ 0.03\\ 
2010-05-02 & 53.3 & 3266 & 0.105 $\pm$ 0.006 & 0.084 $\pm$ 0.005 & 0.021 $\pm$ 0.003 & 14.36 $\pm$ 0.03\\ 
2010-05-06 & 57.0 & 2930 & 0.089 $\pm$ 0.005 & 0.067 $\pm$ 0.005 & 0.022 $\pm$ 0.003 & 14.76 $\pm$ 0.03\\ 
2010-05-18 &  69.5 &  530 & 0.082 $\pm$ 0.014 & 0.056 $\pm$ 0.012 & 0.026 $\pm$ 0.008 & 15.41 $\pm$ 0.03\\
2010-05-26 &  77.4 &  850 & 0.078 $\pm$ 0.011 & 0.065 $\pm$ 0.010 & 0.014 $\pm$ 0.005 & 15.70 $\pm$ 0.02\\
2010-06-03 &  86.0 & 1073 & 0.091 $\pm$ 0.010 & 0.074 $\pm$ 0.009 & 0.017 $\pm$ 0.004 & 16.02 $\pm$ 0.03\\
2010-06-11 &  93.0 &  961 & 0.070 $\pm$ 0.009 & 0.066 $\pm$ 0.009 & $<$0.011         & 16.15 $\pm$ 0.03\\
2010-06-19 & 101.9 &  397 & 0.040 $\pm$ 0.011 & 0.034 $\pm$ 0.011 & $<$0.021         & 16.46 $\pm$ 0.05\\
2010-06-27 & 109.2 & 1236 & 0.046 $\pm$ 0.012 & 0.032 $\pm$ 0.010 & 0.014 $\pm$ 0.007 & 16.54 $\pm$ 0.03\\
2010-07-05 & 117.7 & 1230 & 0.056 $\pm$ 0.008 & 0.045 $\pm$ 0.007 & 0.011 $\pm$ 0.003 & 16.72 $\pm$ 0.03\\
2010-07-13 & 125.9 &  348 & 0.066 $\pm$ 0.016 & 0.050 $\pm$ 0.014 & $<$0.030         & 16.87 $\pm$ 0.06\\
2010-07-21 & 133.1 & 1329 & 0.033 $\pm$ 0.006 & 0.027 $\pm$ 0.005 & 0.006 $\pm$ 0.002 & 16.93 $\pm$ 0.04\\
2010-07-29 & 142.0 & 1140 & 0.033 $\pm$ 0.006 & 0.029 $\pm$ 0.006 & $<$0.008         & 17.15 $\pm$ 0.04\\

\cutinhead{\textit{Suzaku}}
2010-04-09 & 30 & 42200 (FI) & 0.557 $\pm$ 0.004 & 0.300 $\pm$ 0.003 & 0.252 $\pm$ 0.003 & ...\\
                      &       &     (BI)        & 0.479 $\pm$ 0.003\tablenotemark{b} & 0.227 $\pm$ 0.002\tablenotemark{b} & 0.251 $\pm$ 0.002 & ...\\
\enddata
\tablenotetext{a}{Days from 2010 March 10}
\tablenotetext{b}{For \textit{Suzaku} FI chips (XIS0, XIS3), the lower energy bound is 0.4 keV}
\label{swift_lc_table}
\end{deluxetable*}

\subsection{\textit{Suzaku}}
We observed V407 Cyg with {\it Suzaku} (Mitsuda et al. 2006) as a target of opportunity between 2010 April 9 21:32 UT and April 10 19:31 UT (sequence number 905001010). Here we concentrate on the data obtained with the X-ray Imaging Spectrometer (XIS) in the 0.3--10 keV band.  Of the three active units of the XIS, two (XIS0 and XIS3) contain a front-illuminated (FI) CCD chip, while one (XIS1) contains a back-illuminated chip with higher sensitivity to low energy photons.  All three were operated in the full-window, imaging mode, obtaining X-ray event data over the full 19\arcmin\ by 19\arcmin\ field of view every 8 seconds.

After standard screening\footnotemark \footnotetext[9]{see http://heasarc.gsfc.nasa.gov/docs/suzaku/processing/criteria\_xis.~html}, there are 42,200 s of good on-source data with each XIS.  We extracted the source event using a 210\arcsec\ radius circular extraction region centered on the source.  The background was taken from an annular region also centered on the source with outer and inner radii of 410\arcsec\ and 240\arcsec, respectively.  No sources were obvious in the background region.  We then created the response files using the {\tt xisrmfgen} and {\tt xissimarfgen} tools.  Since there were no differences between the XIS0 and XIS3 data, we have combined them in the subsequent analysis.  We ignored data below 0.4 keV for the combined FI spectrum, and below 0.3 keV for the XIS1 spectrum, and grouped the channels so that each had at least 25 source counts (in order to facilitate the use of $\chi^{2}$ statistics in spectral fitting). The XIS1 data appear to have a different normalization, so we have allowed a cross-normalization term in subsequent fitting. 

We inspected the background-subtracted HXD/PIN data to ascertain if V407 Cyg exhibited any excess hard emission beyond the extrapolation of the components detected with the XIS.  For the non-X-ray background, we used the tuned background model, and we used ``typical,'' high-latitude cosmic X-ray background spectrum \citep{Boldt87}.  No signal is detected above 20 keV, with a net count rate of $-9.7 \pm 2.7 \times 10^{-3}$ cts\,s$^{-1}$ in the 20--80 keV range, or $-9.6 \pm 2.4 \times 10^{-3}$ cts\,s$^{-1}$ in the more sensitive 20--50 keV range.  In the 15--20 keV range, an apparent signal of $+5.7 \pm 2.1 \times 10^{-3}$ cts\,s$^{-1}$ was detected, although given the systematic uncertainties in the diffuse background, we consider this a non-detection.
\section{An overview of the outburst evolution}
\label{overview}

\subsection{Optical and UV lightcurves}
The optical and ultraviolet flux from V407 Cyg faded over the entire observation period covered in this work.  The optical V band lightcurve (comprised of verified observations obtained from the AAVSO archive) and the UV data from the \textit{Swift} UVOT are presented in Fig. 1.  V407 Cyg had an observed brightness of $\sim$8.5 magnitudes in the V band, and $\sim$12.6 magnitudes in the UVM2 bandpass at the time of the first \textit{Swift} ToO.  The source then faded in all subsequent observations.  The rate of decline was not constant, and not the same in the optical as in the UV.  From day 5 until day 45 the optical UV emission faded slowly, with a decrease of only one magnitude in the UVM2 filter.  This decay was much slower than that observed in the optical over the same time period, where the V band magnitude decreased from $\sim$8.5 to $\sim$11 (see lower panel of Fig 1) .  After day 45, the UV brightness began to decline much more rapidly, decreasing by one magnitude in just 12 days .  This change in decline rate was also observed in the optical lightcurve, and corresponds to a change in the slope of the X-ray lightcurve at approximately the same time (see Fig 2.).  Finally, around day 60 the rate of decline slowed again, this time fading by one magnitude in approximately twenty days. Again, this change in the decline rate coincided with a similar change in the optical lightcurve, and also with a change in X-ray behavior, in this case the cessation of X-ray fading and transition to approximately constant brightness at day 70.   

\begin{figure}[h]
\begin{center}
\includegraphics[width=3.25in]{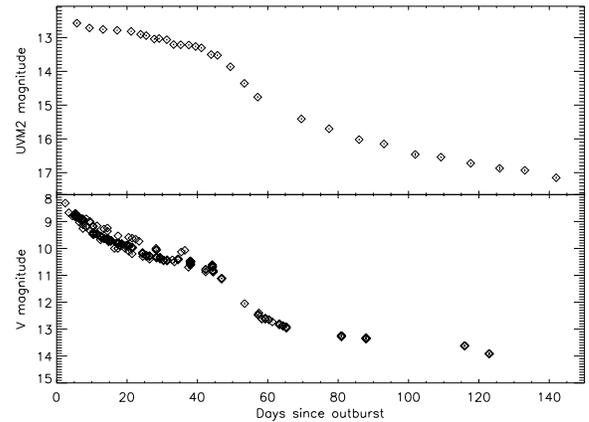}
\caption{\textit{Upper panel:} \textit{Swift} UVOT UVM2 filter light curve.  \textit{Lower panel:} AAVSO V band verified light curve over the same time period.}
\end{center}
\end{figure}
 
\subsection{X-ray lightcurve}
The evolution of the X-ray emission over the course of the outburst was dramatically different from that observed in the optical and UV filters. We present the 0.3--10 keV lightcurve observed with {\it Swift} in the upper panel of Fig. 2, and the hardness ratio (defined as the ratio of the count rates in the 0.3--2 keV and 2--10 keV bands) in the lower panel. V407 Cyg was detected from the first observation onwards, and the source was initially faint (0.011 cts s$^{-1}$ in the 0.3--10 keV range) and rather hard, with twice as many counts above 2 keV as below.  Over the first two weeks of the outburst there was little evolution in either the count rate or the hardness ratio.  
\begin{figure}[h]
\begin{center}
\includegraphics[width=3.25in]{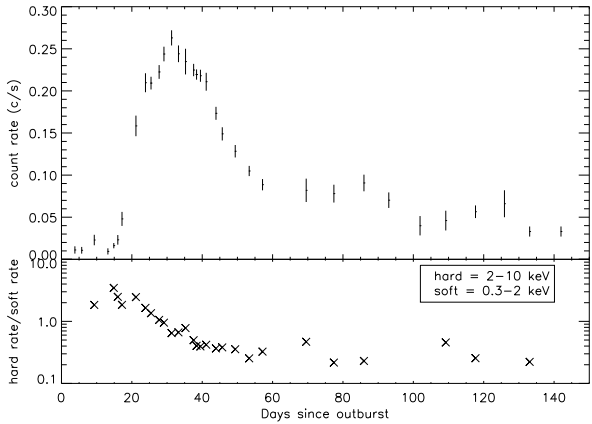}
\caption{{\tt Upper panel:} XRT lightcurve in 0.3--10 keV range over the course of outburst.  {\tt Lower panel:} Hardness ratio evolution with time.  The hardness ratio is defined as the ratio of the 2--10 keV and 0.3--2 keV count rates.}
\end{center}
\end{figure}

Then, at around day 16, the count rate started to rise dramatically, increasing by a factor of 10 from the initial values on day 23.  The system reached a peak count rate of 0.26 cts s$^{-1}$ on day 31 (2010 April 10).  As the source increased in brightness, the hardness ratio declined, indicating that the emission was becoming softer.  After day 31, the emission began to fade, and by day 49 the count rate had decreased by a factor of two from its peak value, reaching 0.12 cts s$^{-1}$.  From this point forward, the hardness ratio remained approximately flat, with a value of $\sim$0.4.  The fading continued until day 70, at which point the 0.3-10 keV count rate remained approximately constant (with some fluctuations) over the subsequent 6 weeks.    

\section{A detailed snapshot obtained at peak X-ray brightness}
\label{spectrum}
The \textit{Suzaku} spectrum, obtained at the peak X-ray brightness of the outburst, offers a highly detailed snapshot of the emission of the system.  The spectrum is dominated by continuum emission, but some lines are also present, most notably those of He-like Si, S and Fe at 1.9, 2.5 and 6.7 keV, respectively.  There also a distinct soft component, peaking at 0.5 keV.  Our goal is to develop a physically motivated model which reproduces the observed spectrum.  Since the emission is clearly complex, we approached this task in stages by focusing on specific features of the spectrum and then adding components as needed.  

To begin, we ignored all obvious emission features and modeled only the continuum.  We included a constant to account for the different normalization of the FI and BI chips, and then tied all model components between the front and back illuminated CCDs.  In all models, we include a {\tt tbabs}\footnotemark \footnotetext[10]{see http://pulsar.sternwarte.uni-erlangen.de/wilms/research/tbabs/} neutral absorber using the abundances of \citet{Wilms00} to represent the interstellar absorption, and fix this value at 2.65 $\times$ 10$^{21}$ cm$^{-2}$.  This value was obtained using the E(B-V) value of 0.5 $\pm$ 0.05 \citep{Shore11} derived from various diffuse interstellar band (DIB) features and the E(B-V) to N(H) conversion relation of \citet{Predehl95}.  The parameters of each model are shown in Table 2, in the "Continuum only" section.  

A single temperature bremsstrahlung component absorbed only by the ISM (model 1 in Table 2) cannot reproduce the data---while the 1--10 keV continuum is fitted reasonably well by this model, large residuals are seen at lower energies.  Adding a second temperature component (model 2) does not improve the fit.  However, the addition of an additional {\tt tbabs} absorber attenuating both plasma components (model 3) results in a dramatically improved fit ($\chi^{2}$/$\nu$ = 1.22 for 1087 degrees of freedom.)  The additional absorber has a column density of 7.6 $\times$ 10$^{21}$ cm$^{-2}$, and the two plasma temperatures are 0.08 and 3.2 keV.  The presence of this additional absorber is reasonable given the dense wind of the Mira companion.  We also tried a blackbody to account for the soft flux (model 4).  The fit is statistically just as good ($\chi^{2}$/$\nu$ = 1.23 for $/nu$ = 1087), and returns similar parameter values for the intrinsic absorber (N(H) = 7.2 $\times$ 10$^{21}$ cm$^{-2}$) and the bremsstrahlung component (kT = 3.3 keV).  The blackbody component has a temperature of 57 eV.

In Figure 3, we show the complete 0.3--10 keV spectrum overplotted with model 4 from Table 2.  The residuals of the model fit are plotted in units of $\Delta\chi^{2}$ in the lower panel of the figure, and clearly demonstrate the presence of the emission lines noted above. The most significant feature in the model residuals is the He-like Fe emission line at 6.7 keV.  Since the analysis of the continuum showed that the X-ray spectrum is comprised of two distinct emission components (dominating the emission above and below 1 keV, respectively,) we consider each separately as we move forward in the development of our spectral model. 

\begin{figure}[h]
\begin{center}
\includegraphics[width=2.25in,angle=270]{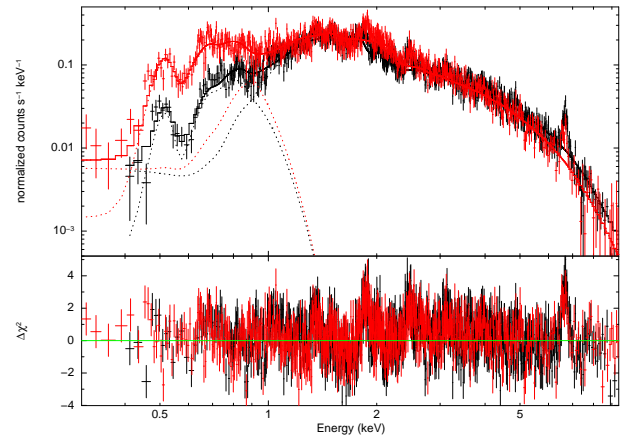}
\caption{\textit{Suzaku} front illuminated XIS0,3 spectrum (\textit{black}) and back-illuminated XIS1 spectrum (\textit{red}) obtained on 2010 April 9, or day 30 of the outburst, shown with an absorbed blackbody + bremsstrahlung model (model 4 in Table 2).  This fit was determined using the continuum only (see Section 4.1).  The fit residuals are shown in the lower panel, clearly demonstrating the existence of emission lines in the data.}
\end{center}
\end{figure} 

\subsection{The hard (1--10 keV) emission}
For the harder component, we consider only the 1--10 keV region of the spectrum.  In order to model the continuum and the emission lines at the same time, we use the {\tt apec} thermal plasma models in Xspec.  These are models of collisional ionization equilibrium (CIE) plasmas developed by \citet{Smith01}, and include contributions to the emission from free-free, recombination and line processes.  The components of these models are presented in Table 2 in the ``Hard X-rays only" section.

We first fit an absorbed single temperature {\tt apec} model to the 1--10 keV data, using the temperature and absorbing column found from the continuum fitting as starting parameter values, and allowing the abundance of the plasma to vary freely.  We again included a fixed value absorber with N(H) = 2.65 $\times$ 10$^{21}$ cm$^{-2}$ in order to represent the interstellar absorption contribution.  The resulting model (model 5 in Table 2) has $\chi^{2}$/$\nu$ = 1.44/1449, and similar parameter values to the bremsstrahlung fit to the continuum (kT = 2.8 $\pm$ 0.1 keV; N(H)$_{int}$ = 6.9 $\pm$ 0.3 $\times$ 10$^{21}$ cm$^{-2}$, where the uncertainties are the 90\% confidence intervals).  The best fit abundance value is 0.36 $\pm$ 0.04 Z$_{\odot}$.  Examining the residuals of this model, the continuum emission and He-like Fe line at 6.7 keV are reproduced very closely.  However, the large ($\sim$4$\sigma$) residuals at the He-like Si and S lines at 1.9 and 2.5 keV that were seen in the simple model fit in Figure 3 remain.

The existence of these strong He-like lines is at odds with the temperature derived for the plasma. In a 2.8 keV (3.4 $\times$ 10$^{7}$ K) plasma less than 20\% of S and no Si at all would be helium-like.  In fact, we expect both elements to be almost completely ionized at this temperature if the gas is in collisional ionization equilibrium \citep[see e.g. ][]{Bryans09}.  We cannot account for the presence of these lines by adding additional temperature components, or multi-temperature plasma models such as {\tt cemekl}. Allowing the abundances to vary does not improve the fit either. This is because the plasma temperature is determined primarily by the fit to the continuum emission,  and appears to be well constrained.  At this temperature, we will never observe He-like lines of Si or S no matter how abundant they are, since these elements are almost completely ionized.

Instead, the strength of these lines may indicate that some fraction of the emitting gas has only recently been shocked, and has not yet had enough time to come into CIE.  If insufficient time has passed for ionization equilibrium to be achieved by particle collisions, then a larger fraction of ions of these species will be in lower ionization stages, increasing the observed strengths of the He-like triplet lines.  In order to test this, we added a non-equilibrium plasma component to the model, implemented in Xspec as {\tt nei}.  This model is characterized by the temperature, abundance and emission measure of the plasma as in the {\tt apec} models, and also by the ionization age $\tau$, defined as the product of the plasma density $n$ and the time since the passage of the shock front, $t$.  This parameter in essence describes how much the plasma has progressed towards coming into collisional ionization equilibrium since it was shocked.  Denser plasmas achieve equilibrium faster since more collisions take place.  We tied the abundance of the {\tt nei} component to that of the {\tt apec} plasma, since it contributes the Fe line at 6.7 keV, and this provides a strong constraint on the abundance of the emitting material.

The addition of the {\tt nei} component improves the fit($\chi^{2}$/$\nu$ = 1.08/1446), and the new model (model 6 in Table 2) accounts for the presence of the He-like lines.  The intrinsic absorber has a higher value in this model, with N(H) = (11.1 $\pm$ 0.5) $\times$ 10$^{21}$ cm$^{-2}$.  The temperature of the {\tt nei} plasma is 3.9$^{+0.5}_{-0.4}$ keV, with $\tau$ = (2.4$^{+0.5}_{-0.3}$) $\times$ 10$^{10}$.   The temperature of the {\tt apec} component is 2.5 $\pm$ 0.1 keV, and the best fit abundance is 0.57 $\pm$ 0.05 Z$_{\odot}$.  Comparing the normalizations of the two components, we find that $\sim$80\% of the flux originates in the {\tt apec} component, i.e. in plasma that is already in ionization equilibrium.  The remaining 20\% is from plasma which is still coming into equilibrium, as indicated by the ionization age of the {\tt nei} component.  As a check, we also tried fitting two {\tt nei} components.  We found that the ionization age of one of the {\tt nei} components tended towards large values, suggesting a plasma that was already in equilibrium.  The other {\tt nei} component had very similar values to the {\tt apec} + {\tt nei} fit.  The presence of a CIE plasma contributing most of the flux appears to be required to account for the bright He-like Fe line observed at 6.7 keV---at the temperatures found by the fit, Fe will not be ionized until $\tau$ $\sim$ 10$^{12}$ cm$^{-3}$ s \citep{Smith10}.

The normalizations of the model components are related to their emission measures, defined as

\begin{equation}
EM = \int_{V} n_{e} n_{i}~ dV  \simeq~ n^{2}V. 
\end{equation}  

Assuming a distance to V407 Cyg of 2.7 kpc, the emission measures of the two model components are 9.9 $\times$ 10$^{56}$ cm$^{-3}$ (the {\tt apec} plasma) and 2.5 $\times$ 10$^{56}$ cm$^{-3}$ (the {\tt nei} plasma).  

\begin{turnpage}
\begin{deluxetable*}{lccccccccccc}
\tabletypesize{\tiny}
\tablecaption{Model fits to {\it Suzaku} data} 
\tablewidth{0pt}
\tablehead{
\colhead{Model} & \colhead{N(H)\tablenotemark{1}$_{ISM}$}  & \colhead{N(H)$_{int}$}  & \colhead{$kT_{1}$} &  \colhead{log(norm$_{1}$)} &   \colhead{$kT_{2}$} & \colhead{log(norm$_{2}$)} & \colhead{$kT_{3}$} & \colhead{log(norm$_{3}$)} & \colhead{$Z$} & \colhead{$\tau$} &  \colhead{$\chi^{2}_{\nu}$}  \\
& \colhead{(10$^{21}$ cm$^{-2}$)} & \colhead{(10$^{21}$ cm$^{-2}$)} &  \colhead{(keV)} & & \colhead{(keV)} &  & \colhead{(keV)} &  & \colhead{($Z_{\odot}$)} & \colhead{(10$^{10}$ cm$^{-3}$ s)} & \\ 
}
\startdata   
\cutinhead{\textit{Continuum Only}}
1) tb$_{ISM}$*br$_{1}$  & 2.65                    & \nodata & 5.8 $\pm$ 0.2  & -2.62 $\pm$ 0.01 & \nodata & \nodata & \nodata & \nodata & \nodata  & \nodata  & 2728/1090 \\
2) tb$_{ISM}$*(br$_{1}$+br$_{2}$) &  2.65 & \nodata & 6.0 $\pm$ 0.2 & -2.63 $\pm$ 0.01  & 0.11 $\pm$ 0.01 &  -0.73$^{+0.06}_{-0.05}$ & \nodata  & \nodata & \nodata  & \nodata & 2496/1088 \\
3) tb$_{ISM}$*tb$_{int}$*(br$_{1}$+br$_{2}$) & 2.65 & 7.6$^{+0.2}_{-0.3}$ & 3.3 $\pm$ 0.1 & -2.37 $\pm$ 0.03 & 0.076 $\pm$ 0.002 & 2.29$^{+0.05}_{-0.15}$ & \nodata & \nodata & \nodata & \nodata  & 1326/1087 \\
4) tb$_{ISM}$*tb$_{int}$*(br$_{1}$+bb$_{2}$) & 2.65 & 7.2$^{+0.4}_{-0.3}$ & 3.3 $\pm$ 0.1 & -2.39 $\pm$ 0.01 & 0.06$^{+0.02}_{-0.01}$ & -1.27$^{+0.14}_{-0.09}$  & \nodata & \nodata & \nodata & \nodata  & 1335/1087 \\
\cutinhead{\textit{Hard X-rays only}}
5) tb$_{ISM}$*tb$_{int}$*ap$_{1}$ & 2.65 & 7.2 $\pm$ 0.3 & 2.8 $\pm$ 0.1 & -1.82$^{+0.01}_{-0.02}$ & \nodata & \nodata & \nodata & \nodata & 0.36 $\pm$ 0.04 & \nodata & 2091/1449 \\
6) tb$_{ISM}$*tb$_{int}$*(ap$_{1}$+nei$_{2}$) & 2.65 & 11.1 $\pm$ 0.5 & 2.5 $\pm$ 0.1 &  -1.92 $\pm$ 0.01 & 3.9 $\pm$ 0.5 & -2.52$^{+0.04}_{-0.05}$ & \nodata & \nodata & 0.57 $\pm$ 0.05 & 2.4$^{+0.5}_{-0.3}$ & 1555/1446 \\
\cutinhead{{\it Complete Spectrum}}
7) tb$_{ISM}$*tb$_{int}$*(ap$_{1}$+nei$_{2}$+br$_{3}$) & 2.65 & 15.6 $\pm$ 0.1 & 2.3 $\pm$ 0.1 & -1.74 $\pm$ 0.01 & 3.9$^{+0.4}_{-0.3}$ & -2.52$^{+0.04}_{-0.05}$ & 0.050 $\pm$ 0.001 & 6.15$^{+0.02}_{-0.05}$ & 0.81$^{+0.06}_{-0.05}$ & 2.2 $\pm$ 0.2  & 1832/1657 \\
8) tb$_{ISM}$*tb$_{int}$*(ap$_{1}$+nei$_{2}$+ap$_{3}$) & 2.65 & 17.3$^{+0.4}_{-0.3}$ & 2.1 $\pm$ 0.1 &  -1.92$^{+0.01}_{-0.02}$ & 4.2$^{+0.4}_{-0.3}$ & -2.38$^{+0.01}_{-0.05}$ & 0.040 $\pm$ 0.001 & 5.57$^{+0.04}_{-0.09}$ & 0.90$^{+0.07}_{-0.06}$ & 2.1 $\pm$ 0.2  & 1981/1657 \\
9) tb$_{ISM}$*tb$_{int}$*(ap$_{1}$+nei$_{2}$+bb$_{3}$) & 2.65 & 15.4$^{+0.3}_{-0.4}$ & 2.3 $\pm$ 0.1 & -1.74 $\pm$ 0.01 & 3.9$^{+0.4}_{-0.3}$ & -2.52$^{+0.04}_{-0.05}$ & 0.038 $\pm$ 0.001 & 2.06$^{+0.15}_{-0.19}$  &  0.80$^{+0.06}_{-0.05}$ & 2.2 $\pm$ 0.2  & 1836/1657 \\
\enddata 
\label{suzaku_fits}
\tablenotetext{1}{Fixed to value derived from optical spectroscopy by Shore et al. (2011).}
\tablenotetext{2}{norm$_{brem}$ = $\frac{3.02 \times 10^{-15}}{4\pi D^{2}} \int n_{e} n_{i} dV$, where $D$ is the distance to the source in units of cm and $n_{e}$ and $n_{i}$ are the electron and ion number densities, respectively, in units of cm$^{-3}$.}
\tablenotetext{3}{norm$_{apec}$, norm$_{nei}$ = $\frac{10^{-14}}{4\pi D^{2}} \int n_{e} n_{i} dV$, where the parameters have the same meaning as in the bremsstrahlung model. }
\tablenotetext{4}{norm$_{BB}$ = $\frac{L_{39}}{D_{10}^{2}}$, where $L_{39}$ is the bolometric luminosity of the source in units of 10$^{39}$ erg s$^{-1}$, and $D_{10}$ is the distance to the source in units of 10 kpc.}
\end{deluxetable*}
\end{turnpage}     

\subsection{Fitting the entire spectrum}

With a satisfactory model for the hard emission, we now expand our analysis to fit the entire 0.3--10 keV spectrum.  We learned from the continuum-only fits that the soft component could be modeled equally well by thermal bremsstrahlung emission with kT = 0.08 keV or a blackbody with temperature 57 eV.  In either case, the soft component is subject to the same high line of sight absorption as the 1--10 keV emission.  We will explore both of these possibilities for the soft emission by adding to the model developed in Section 4.1.  Summaries of the resulting models are presented in Table 2 in the ``Complete Spectrum" section.

We find a reasonable fit to the data ($\chi^{2}$/$\nu$ = 1.11/1657) fitting the softest emission with a thermal bremsstrahlung model (model 7 in Table 2.)  The parameters of the {\tt apec} and {\tt nei} components remain largely unchanged, although the model fit implies a higher value of the metallicity (0.8 $Z_{\odot}$) than before.  The bremsstrahlung component has a temperature of 0.05 keV, slightly lower than that found for the continuum fit, and a normalization of 1.4 $\times$ 10$^{6}$,  corresponding to an emission measure of  $\sim$4 $\times$ 10$^{64}$ cm$^{-3}$.  The most likely origin for such low temperature plasma in this system is the reverse shocked ejecta.  However, the high emission measure seems to rule out this possibility, as it implies a very high ejecta mass, $M_{ej}$. 

To estimate $M_{ej}$, we assume that the ejecta are comprised only of hydrogen, and that they have been completely ionized by the reverse shock, which has already reached the white dwarf surface. We further assume that the shocked ejecta occupy a roughly spherical volume of radius $R = v_{ej}t$, where $v_{ej}$ is the initial ejecta velocity and $t$ is the time since outburst. Taking an ejecta expansion velocity of 3200 km s$^{-1}$ for 30 days (based on the FWZI of the Balmer lines, see Munari et al. 2011 and Shore et al. 2011), then the emitting volume is $\sim$2 $\times$ 10$^{45}$ cm$^{3}$.  Inserting this volume estimate into Equation 1, we obtain an estimate of the ejected mass of $\sim$7 $\times$ 10$^{-3}$ M$_{\odot}$.  This mass estimate can be lowered somewhat by accounting for deceleration of the ejecta, which would reduce the volume.  However, even considering a volume 10 times smaller than the one assumed here only reduces the mass estimate to 2 $\times$ 10$^{-3}$ M$_{\odot}$.  Such high ejected masses would be unprecedented in a nova outburst, and appears to be at odds with the fast evolution of the optical lightcurve.

To allow for the possibility of more complex soft emission, we fit the data using an {\tt apec} model (model 8).  Although the resulting emission measure is smaller, it is still extremely large (a few 10$^{63}$), implying an ejected mass of at least a few 10$^{-4}$ M$_{\odot}$.  Furthermore, the fit is not as good ($\chi^{2}$/$\nu$ = 1.20/1657), and clear residuals are seen in the data at the location of spectral lines of e.g. O VIII Ly $\alpha$ at $\sim$0.8 keV.  We note that Shore et al. achieved acceptable fits to the soft component observed in the summed {\it Swift} spectrum using a variable abundance {\tt apec} model with extremely super-solar N, O and Ne abundances.  We tried using a similar model to fit the soft component in the {\it Suzaku} data, allowing the abundances of N, O and Ne in the low temperature {\tt apec} component to vary.  However, we found no combination of parameters resulting in $\chi^{2}$/$\nu$ $<$ 1.4, and large residuals are still apparent in the data, particularly between 0.7 and 1 keV.  The emission measure required to reproduce the observed flux in this model is lower, although it still implies a high ejecta mass at odds with the fast optical evolution of the outburst. This contradiction, in combination with the extremely high abundances required and the poorer fit to the data, lead us to conclude that a thermal, line emitting plasma is an unlikely origin of the soft emission.

Finally, we considered a blackbody origin for the soft emission (model 9 in Table 2).  Including such a source is reasonable if we are seeing the supersoft emission from the still burning white dwarf.  Although the intrinsic absorption of the red giant wind is high,  it is feasible that the supersoft emission is observable due to its high luminosity (typically $>$ 10$^{37}$ erg s$^{-1}$).  The fit found for this model is good, with $\chi^{2}$/$\nu$ = 1.11/1661.  Again, the parameters for the hard components remain largely unchanged from the 1--10 keV only fits.  The temperature of the blackbody is 38 eV, corresponding to $\sim$440,000 K.  The normalization corresponds to a bolometric luminosity of 8.4 $\times$ 10$^{39}$ erg s$^{-1}$.  While this is unrealistically large, it is now well known from X-ray studies of novae during the supersoft phase that blackbody fits tend to under-predict the temperature and overestimate the luminosity of these shell burning white dwarfs (see e.g \citealt{Osborne11} (RS Oph) and \citealt{Page10} (V2491 Cyg)).  The data and this best fit model are presented in Figure 4.  The 0.3--10 keV flux of the hard component is 9.1 $\times$ 10$^{-12}$ erg s$^{-1}$ cm$^{-1}$, and the luminosity (corrected for absorption) in the same energy range is 2.8 $\times$ 10$^{34}$ erg s$^{-1}$.
\begin{figure}[h]
\begin{center}
\includegraphics[width=2.25in,angle=270]{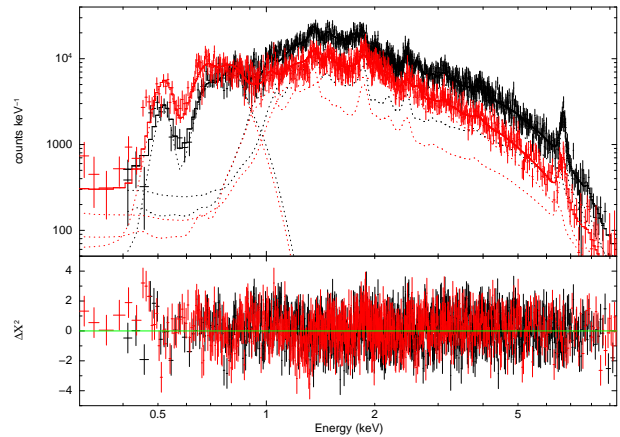}
\caption{XIS 0.3--10 keV spectrum with best fit model 9 from Table 2.  This model has a blackbody for the soft component.  The lower panel shows the fit residuals in units of sigma. The emission line residuals seen in Figure 3 are now gone.}
\end{center}
\end{figure}

We made one further refinement to this model.  Since the intrinsic absorption likely arises in the wind, we tied the abundances of this absorber to that of the {\tt apec} component, which itself originates in the wind shock heated by the ejecta.  The fit is statistically the same as the solar abundance absorber ($\chi^{2}$/$\nu$ = 1.11 for $\nu$ = 1661), and all model parameters agree within the uncertainties, suggesting that the fit is insensitive to the abundance of the material, which is not surprising given the large column density attenuating the soft flux.

As an additional check, we extrapolated our best fit model into the HXD energies.  Using PIMMS v4.3, we find that the model predicts a signal of only 2.9 $\times$ 10$^{-4}$ cts s$^{-1}$ in the 15--20 keV range.  This is consistent with the non-detection in the HXD detector.

\section{Evolution of the two X-ray emitting components over the outburst}

The model we have developed for the \textit{Suzaku} data gives a detailed snapshot of the interaction of the nova ejecta and the Mira wind.  We now utilize that model to interpret the {\it Swift} lightcurve, focusing on the evolution of the two distinct spectral components observed in the {\it Suzaku} spectrum.  As we discussed in Section 3.1, most of the \textit{Swift} observations were too short to collect enough counts for spectroscopy.  However, we can increase the statistics by summing neighboring observations together in order to investigate the origins of the changes in the X-ray count rate and hardness ratio we observed in the lightcurve. 

During the first two weeks of the outburst the count rate remained low with a relatively constant hardness ratio, ranging between 2 and 3.  Since we observed very little evolution in the lightcurve over this period, we summed the first seven observations (up to March 27) together. From this date and forward, we summed three consecutive observations at a time to derive spectral information (e.g., from March 27 to April 2, April 4 to April 8, and so on). If the combined spectrum had less than $\sim$200 counts, we fit our model to the unbinned data and utilized the Cash statistic to assess confidence intervals for our parameters of interest.  If the resulting spectrum had more counts,  we binned the data to give a minimum of 20 counts per bin, and adopted the $\chi^{2}$ statistic to assess best fit parameter values and associated confidence intervals. Hereafter, when we mention the number of days after the $\gamma$--ray trigger, we refer to the time of the middle observation in each summed group of three. 

The {\tt nei} component in our best fit {\it Suzaku} model requires good statistics in the emission lines in order to pinpoint the ionization age and the contribution of this component versus the CIE plasma.  Since the  {\it Swift} spectra have rather low numbers of counts (even after summing consecutive observations), and since the {\tt nei} component only contributes $\sim$20\% of the flux at the time of peak brightness, we decided to use a single {\tt apec} model only to approximate the hard flux.  Therefore, our {\it Swift} model takes the form {\tt tbabs*tbabs*(bb+apec)}.  We fixed the abundance of the {\tt apec} component at 0.8 $Z_{\odot}$, the N(H) of the first absorber to 2.65 $\times$ 10$^{21}$ cm$^{-2}$ (again to represent the ISM contribution), and allowed the intrinsic absorption, temperatures and normalizations of the two components to vary.  Furthermore, we fixed the abundance of the intrinsic absorber to that of the {\tt apec} component, since both the emission and absorption arise in the Mira wind.

The {\tt bb+apec} model provides a reasonable fit to the data for all summed spectra after day 17.  The resulting parameter values for the two emitting components, the total flux and contribution from each, and the column density of the intrinsic absorber are plotted in Figure 5.   The model does not, however, result in a good fit to the summed spectrum obtained for the first two weeks of the observation (see Figure 6).  Although the 2--10 keV flux in this spectrum is reproduced well, the model underestimates the counts at E $<$ 1 keV.  There is a large residual around 0.4 keV, which seems to arise primarily in the earliest observations.  It appears that a single blackbody cannot adequately account for the flux between 0.3 and 1 keV.  Furthermore, the normalization of the best fit blackbody returned by the model fitting implies a luminosity in excess of 10$^{43}$ erg s$^{-1}$---a non-physical quantity for this system.  Although this may be an affect of averaging over a relatively large period of time (two weeks), there are insufficient counts to draw any strong conclusions about the poor fit.

\begin{figure*}
\begin{center}
\includegraphics[width=2.3in]{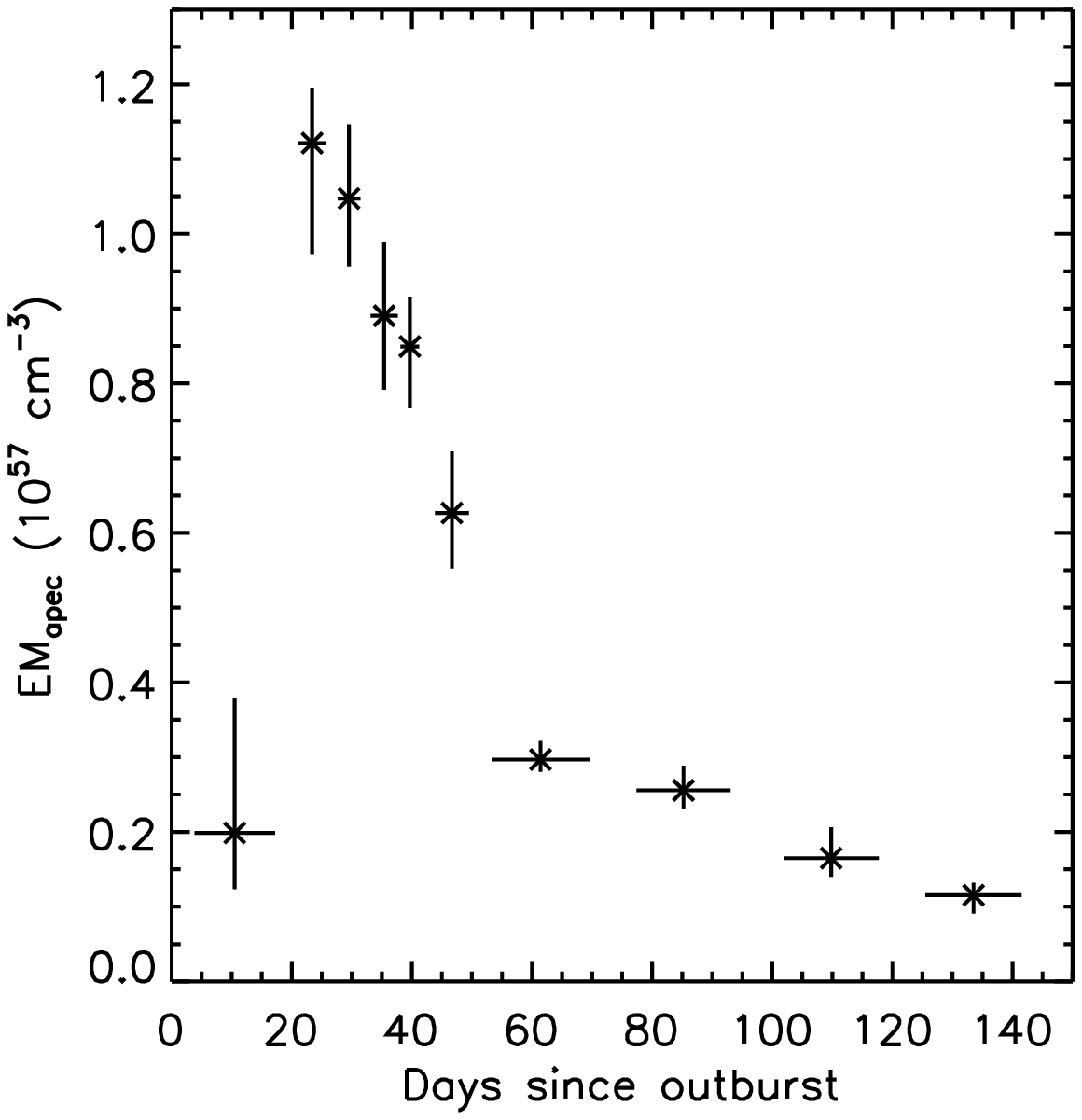} 
\includegraphics[width=2.3in]{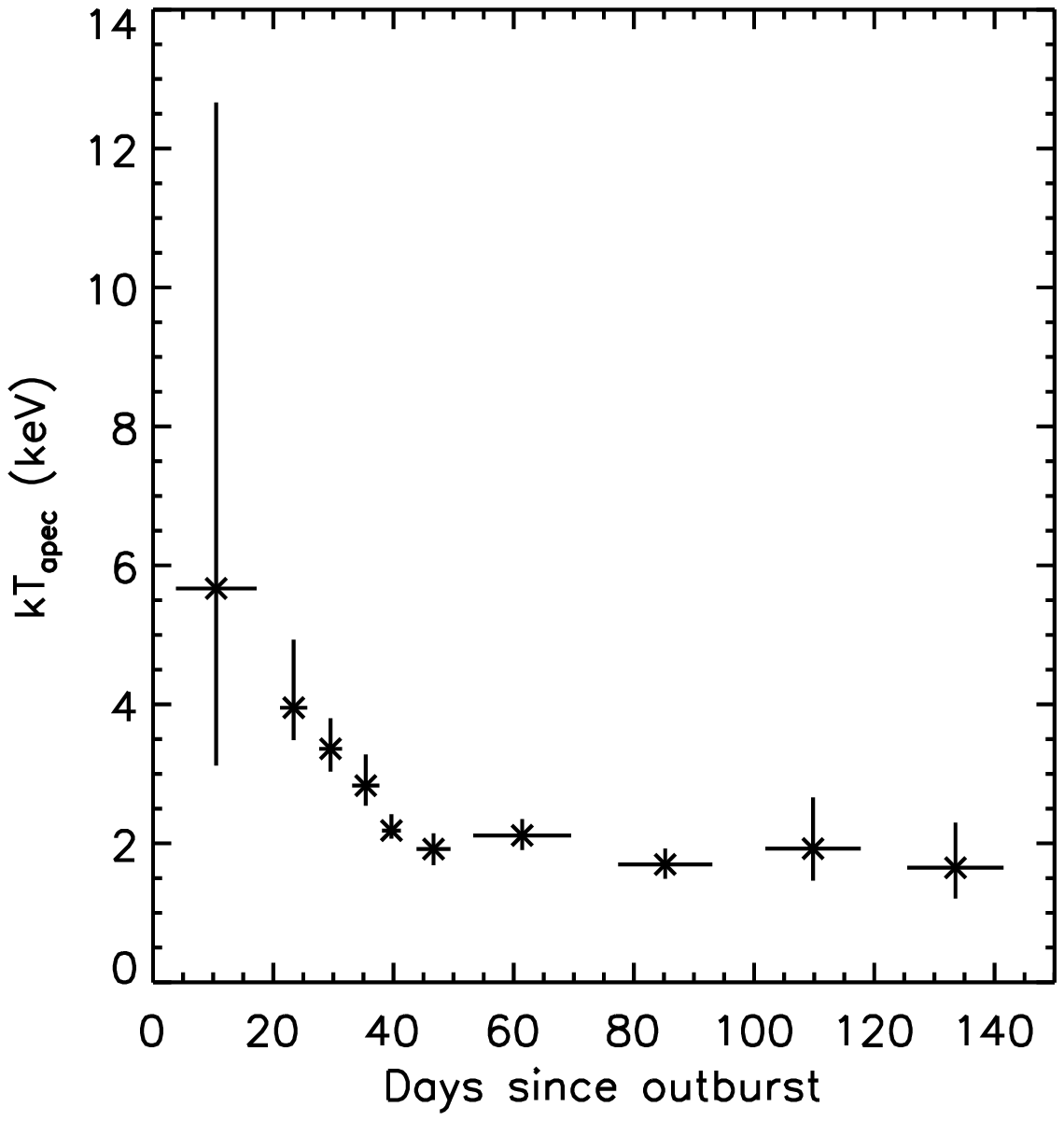}
\includegraphics[width=2.3in]{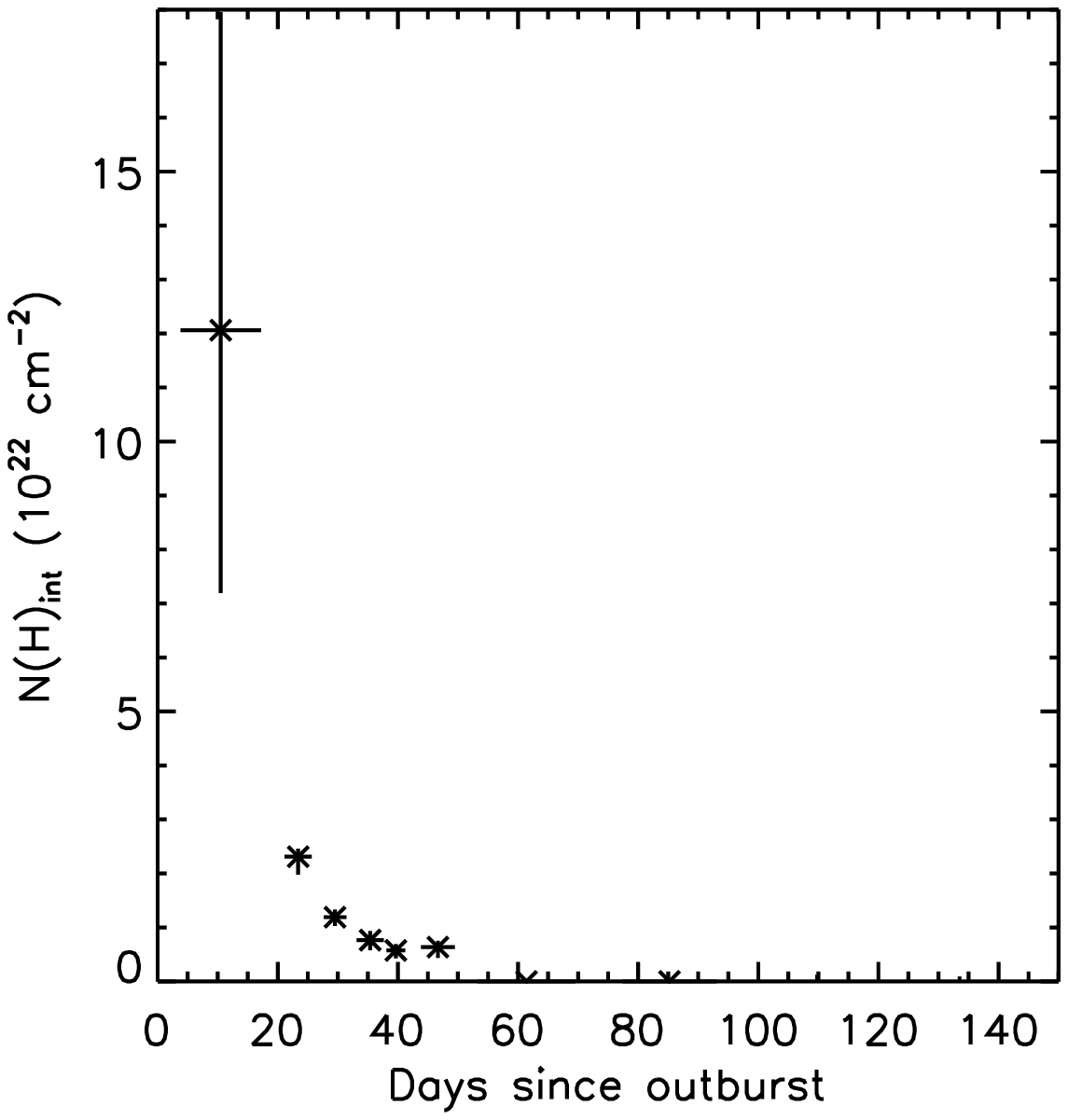} \\
\includegraphics[width=2.3in]{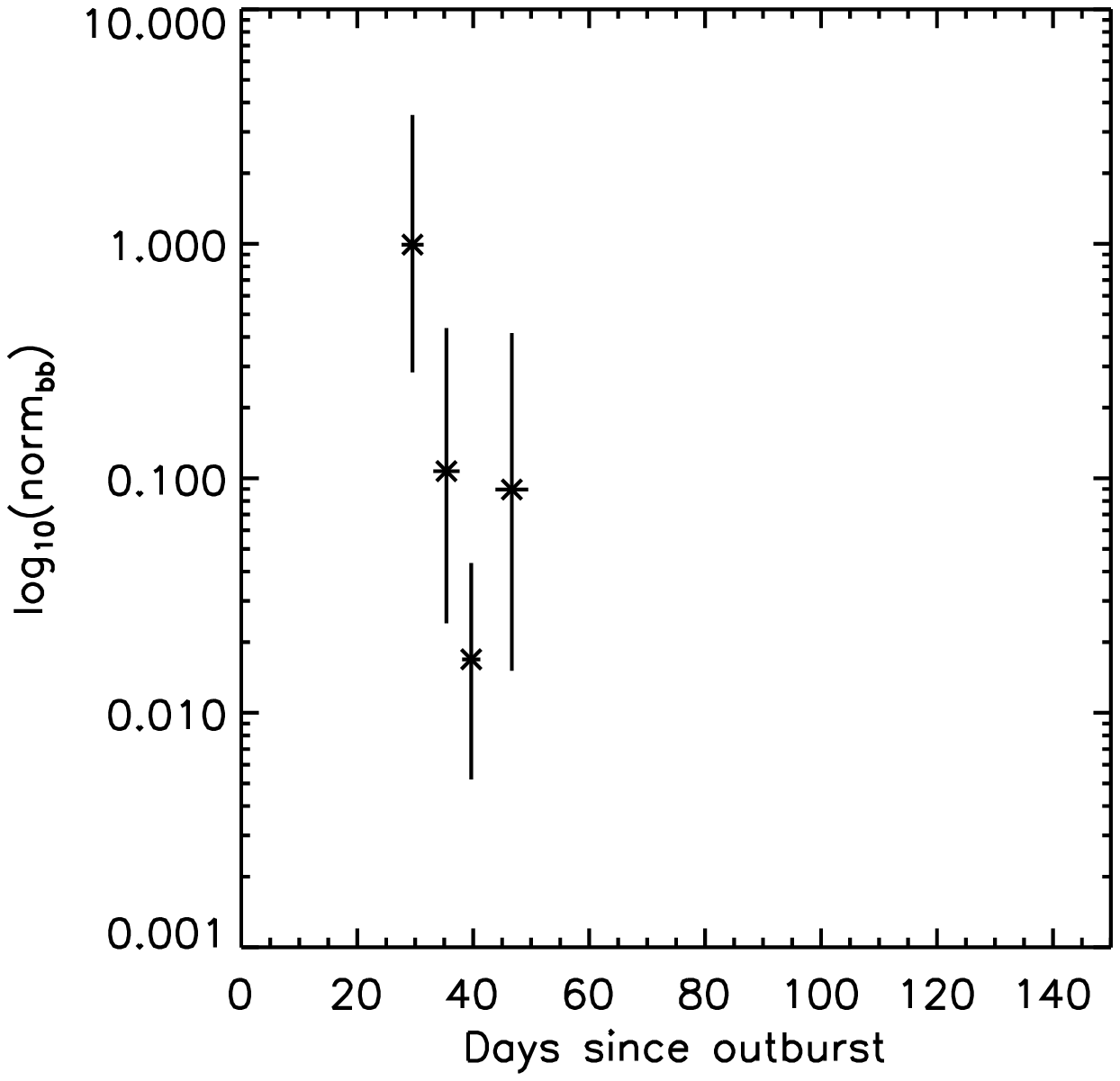} 
\includegraphics[width=2.3in]{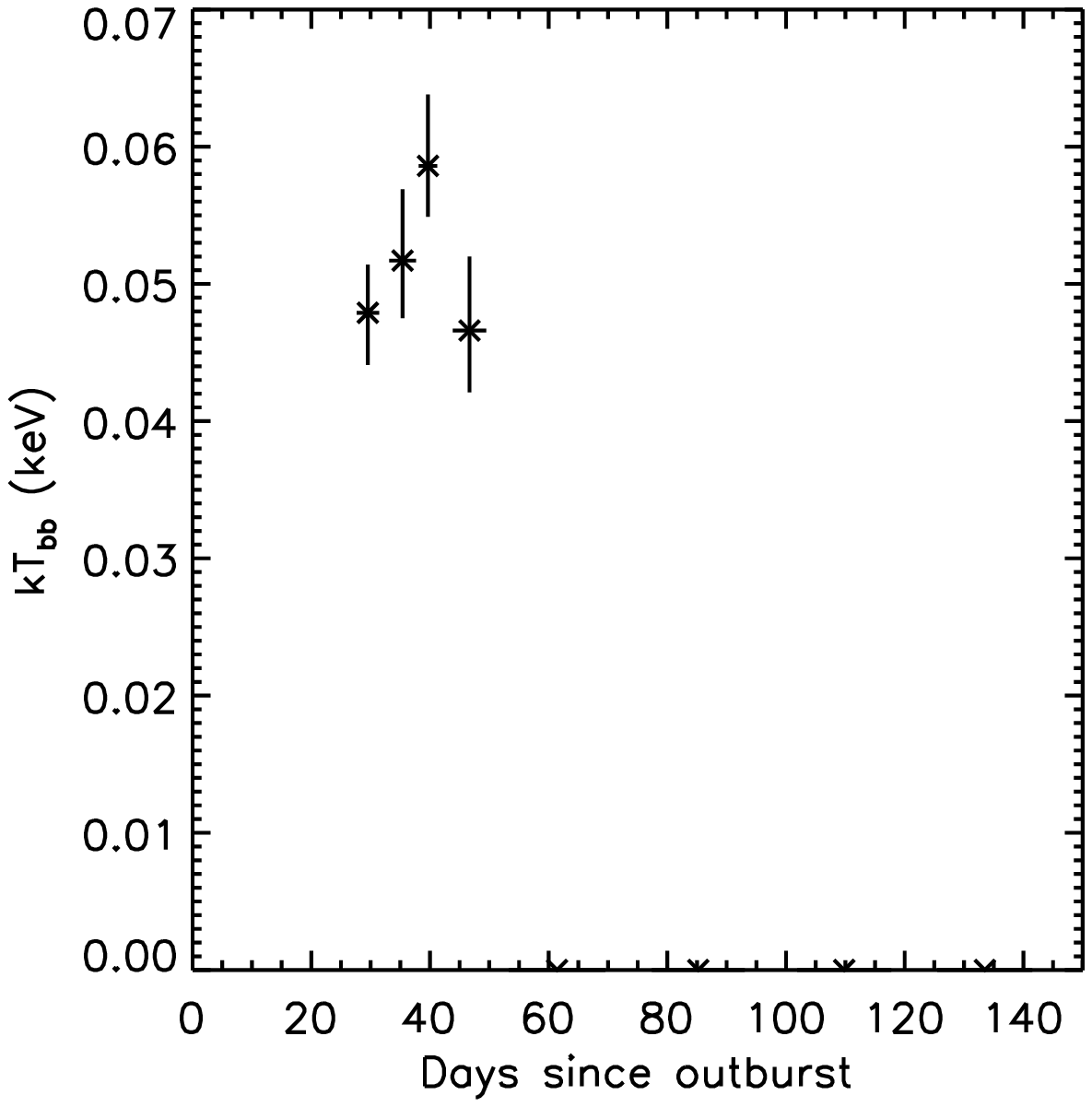}
\includegraphics[width=2.3in]{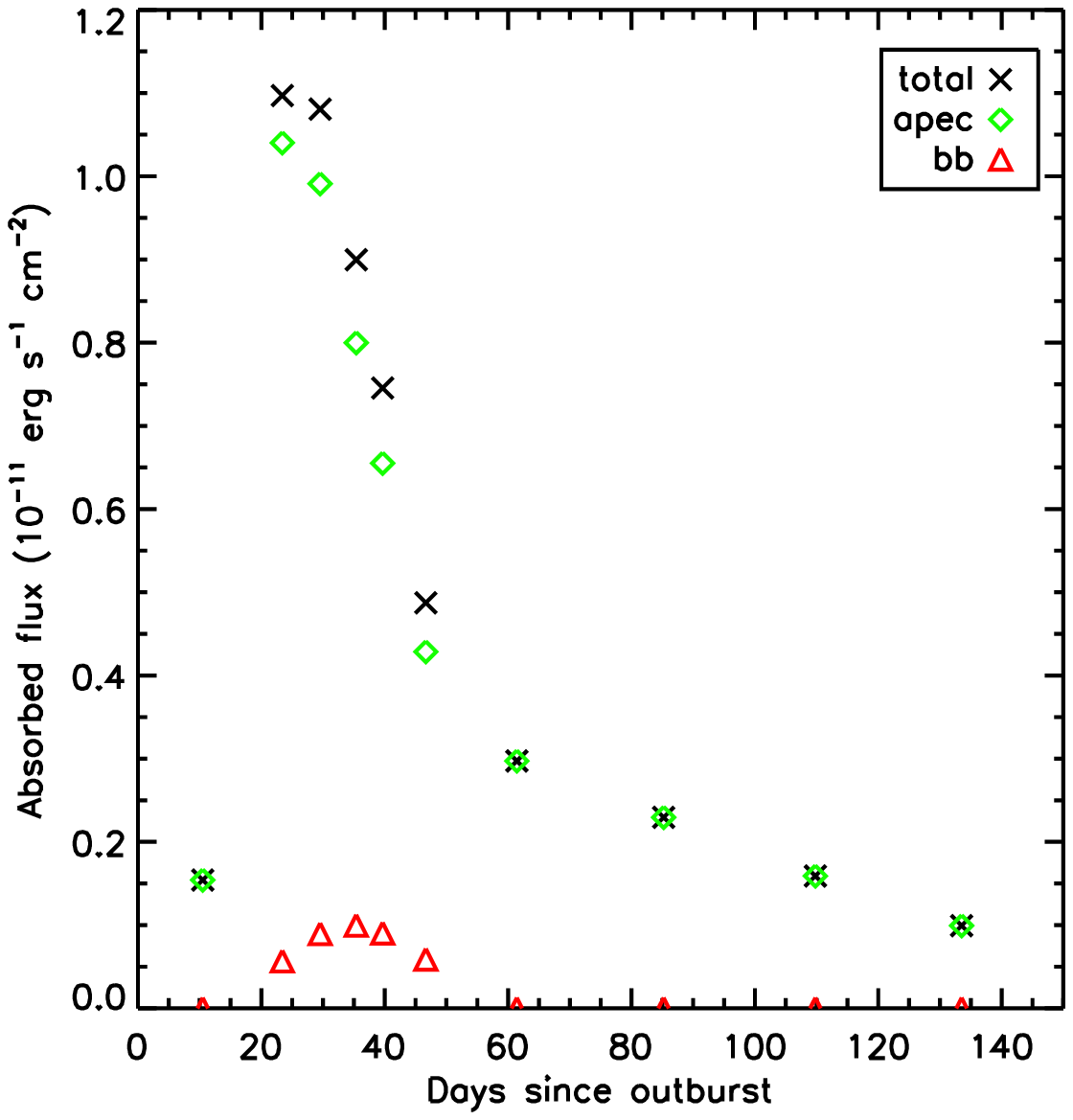}

\caption{Evolution of the absorbed {\tt blackbody + apec} model fit to the {\it Swift} data.  The parameters were determined for 3 day averaged spectra.  {\it Top row:} Evolution of {\tt apec} emission measure (left) and temperature (center), assuming a distance of 2.7 kpc.  Evolution of the column density of the intrinsic absorber (right).  {\it Bottom row:} Evolution of the blackbody normalization and temperature (left and center).  Evolution of the absorbed flux of the total model fit (black crosses), and the contribution from each component, {\tt apec} in green diamonds and {\tt bb} in red triangles (right). } 
\label{fig5} 
\end{center} 
\end{figure*} 

The parameters of the spectral fits show that the shocked gas cools significantly from its highest point during the first two weeks, when the recorded temperature was $\sim$6 keV, to the beginning of the plateau around day 60, when the temperature decreased to $\sim$1.5 keV. After this date, the temperature remains approximately constant.  The emission measure of this component peaks on day 30 at a maximum value of 1.15 $\times$ 10$^{57}$ cm$^{-3}$, in good agreement with the {\it Suzaku} model value (1.24 $\times$ 10$^{57}$ cm$^{-3}$.)  The emission measure then drops by almost an order of magnitude over the subsequent 30 days, coinciding with the drop in temperature.  The decline becomes shallower after day 50, decreasing by only a factor of 3 by day 140.

\begin{figure}[h]
\begin{center}
\includegraphics[width=2.25in,angle=270]{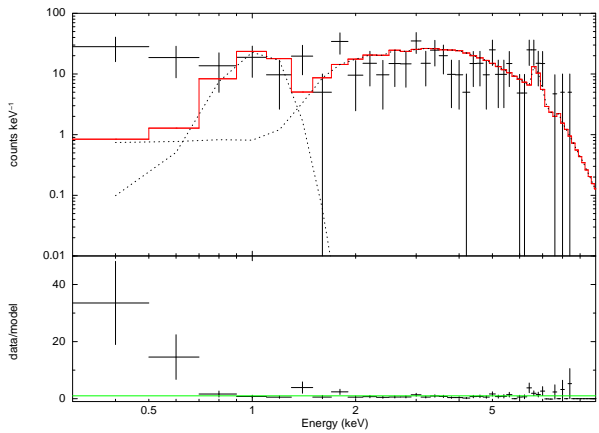}
\caption{Best fit of the {\tt bb+apec} model to the summed data from the first two weeks of the outburst.  The red line is the combined model, and the dashed lines indicate the blackbody and plasma components. The data have been rebinned for plotting purposes to have a minimum of 5 counts per bin.}
\end{center}
\end{figure} 

The blackbody component is first detected after day 20, although the normalization is poorly constrained until day 30.  Between days 20 and 40, there is an apparent increase in the temperature, from 0.03 to 0.06 keV, followed by a drop around day 50 to 0.045 keV.  Over this time period, the normalization varies between 0.02 and 1.5, although with large uncertainty, implying bolometric luminosities in the range 10$^{36}$ to 2.7 $\times$10$^{37}$  erg s$^{-1}$.  As a check, we modeled the individual {\it Swift} spectrum obtained on day 29, just prior to the {\it Suzaku} observation.  We find that the blackbody temperatures agree closely.  The blackbody normalization is about an order of magnitude smaller in the {\it Swift} best fit model than in the {\it Suzaku} model, although within the rather large uncertainties they agree.  The differences are likely due to the sensitivity of the blackbody flux to the intrinsic absorption in the model---small changes in N(H) correspond to large changes in blackbody normalization.  After day 50, the blackbody component is no longer required in the model fit, and its normalization becomes consistent with zero in all subsequent models.  The column density of the intrinsic absorber shows a clear downwards trend with time, from a maximum value of $\sim$3.2 $\times$ 10$^{22}$ cm$^{-2}$ to zero around day 60.  In model fits after this date, the absorption is consistent with an origin in the ISM only.   We note that as the intrinsic absorber decreases in strength, the soft emission would become easier to observe if present at the same flux level, and so the turn off of the blackbody component appears to be real.

In the lower right of Figure 5, we plot the observed fluxes in the 0.3-10 keV energy range of the total model (in black), the harder apec (in green) and the blackbody (in red).  These have not been corrected for absorption.  The majority of the flux at all stages of the evolution (and all flux after day 60) originates in the hard component.  The black body can be seen to brighten and then turn off.  This allows us to understand the hardness ratio evolution shown in Figure 1---the evolution towards a softer spectral state is only marginally influenced by the disappearance of the black body component, and is primarily due to the drop in temperature of the hard component.

\section{A simple model of the nova ejecta---Mira wind interaction}
In light of the X-ray evolution observed in RS Oph during its 2006 outburst, we propose that the hard X-ray component observed in V407 Cyg also originates in the forward shock being driven into the AGB star's wind by the nova ejecta. In order to understand the key features of this evolution---the initial low count rate, the dramatic rise in brightness around day 20 and the subsequent fading---we have developed a simple model of the interaction of the nova ejecta with their environment.  The model tracks the sweeping up of the Mira wind by the ejecta as they travel outwards from the white dwarf, and estimates the key properties of the resulting X-ray emission: the total emission measure, the weighted average X-ray temperature, and the total luminosity.  These can then be compared to the values obtained from the data.  

In spherically symmetric models, where the nova outburst happens at the center of the companion wind, the X-ray quantities of interest  depend only on the distance from the white dwarf since the density of the material encountered by the ejecta is the same in all directions.  This allows the use of simple blast wave approximations, as was done for RS Oph \citep[e.g.][]{Sokoloski06}.  However, if the white dwarf is offset significantly from the origin of the giant companion's wind, these relationships no longer hold since the ambient density depends strongly on the direction of motion of the ejecta.  The ejecta encounter a medium of {\it increasing} density towards the Mira companion,  compared with one of {\it decreasing} density in the opposite direction.

\begin{figure}[h]
\begin{center}
\includegraphics[width=3in]{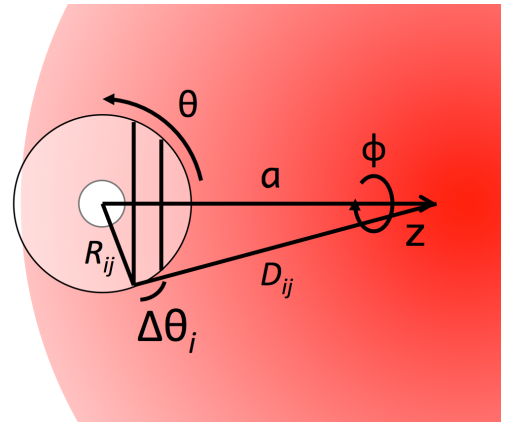}
\caption{The geometry of the system used in the model.  The co-ordinate system is centered at the white dwarf (the white circle), embedded within the wind of the red giant, which is located a distance $a$ away along the $z$ axis.  The ejecta travel outwards from the white dwarf.  We show one example here of a ring of ejecta, with extend $\Delta\theta_{i}$.  The discontinuity surface between the ejecta and the giant wind is located a distance $R_{ij}$ from the white dwarf, and $D_{ij}$ from the giant.}
\end{center}
\end{figure} 

To account for this asymmetry in the environmental density, we utilize the geometry presented in \citet{Abdo10} and illustrated in Figure 7.  The circumbinary environment is described using a spherical coordinate system centered on the white dwarf, with the red giant located a distance $a$ away on the polar axis (where $a$ is the binary separation).   We assume that the nova ejecta are instantaneously ejected as a spherically symmetric shell, which we divide into a series of $i$ rings in polar angle, each centered on and perpendicular to the polar axis joining the two stars. In the $j^{th}$ time step of the model run, the $i^{th}$ ring is characterized by its radial distance from the white dwarf, $R_{ij}$, and its polar angle $\theta_{ij}$ from the line joining the two stars. We also take $R_{ij}$ to specify the shock discontinuity surface for each ring.  Since all points on a given ring are equidistant from the Mira, they are embedded in ambient medium with the same density, which we denote $\rho_{ij}$. For each ring, $\theta_{ij}$ will remain constant in time as we assume radial propagation of the ejecta away from the white dwarf into a medium essentially at rest, and so we denote it $\theta_{i}$ moving forward. Finally, because $\rho_{ij}$ is the same for all points on a given ring, the rings will remain circular as they propagate outwards.

In our model, we consider $i$ = 180 equally spaced rings, each with extent $\Delta\theta_{i}$ = 1$^{\circ}$. The fraction of the total ejected mass in each ring, $M_{i}$, is simply the fraction of the total volume of the ejected shell that each ring occupies at the beginning of the simulation (when the shock wave is spherical),  and is given by

\begin{equation}
M_{i} =  M_{ej} \int^{\theta_{2}}_{\theta_{1}} sin(\theta)  d\theta,
\end{equation}  where $\theta_{1}$ and $\theta_{2}$ are the two angles defining the angular extent of the $i^{th}$ ring.  The ejecta start at the white dwarf surface, and are given an initial velocity, $v_{0}$ = 3200 km s$^{-1}$, based on the FWZI of the hydrogen Balmer lines observed in the earliest phases of the outburst \citep{Munari11,Shore11}.   Therefore, each ring has an initial reservoir of kinetic energy $KE_{i} = 0.5 M_{i} v_{0}^{2}$.

Assuming that the Mira wind can be described by an $r^{-2}$ density profile, the ambient density encountered by the $i^{th}$ ring in the $j^{th}$ time step, $\rho_{ij}$ can be expressed as
\begin{equation}
\rho_{ij} = \rho_{0} \left( \frac{R_{RG}}{D_{ij}} \right )^{2}.
\end{equation} where $R_{RG}$ is the photospheric radius of the red giant, $\rho_{0}$ is the density of the Mira wind at $R_{RG}$, and $D_{ij}$ is the radial distance of the $i^{th}$ ring from the Mira in the $j^{th}$ time step.  $D_{ij}$ is given by \begin{equation}
D_{ij} = \sqrt{a^{2}+R_{ij}^{2}-2aR_{ij}cos(\theta_{i})}-R_{RG},
\end{equation}recalling that $R_{ij}$ is the radial distance of the $i^{th}$ ring from the white dwarf.  The density at the base of the Mira wind, $\rho_{0}$,  is  

\begin{equation}
\rho_{0} = \frac{\dot{M}_{RG}}{4 \pi R_{RG}^{2} v_{RG}},
\end{equation} where $\dot{M}_{RG}$ and $v_{RG}$ are the mass loss rate and terminal velocity of the red giant wind.  The quantities $\dot{M}_{RG}$ and $R_{RG}$ are not well known for V407 Cyg, and so we take fiducial values of $\dot{M}_{RG}$ = 10$^{-7}$ M$_{\odot}$ yr$^{-1}$ and $R_{RG}$ = 500 R$_{\odot}$.  We set $v_{RG}$ = 30 km s$^{-1}$ based on the equivalent width of the Na I D lines at 5890 \AA~ observed both in quiescence \citep{Tatarnikova03b} and outburst \citep{Shore11}.  These values result in $\rho_{0}$ = 4.2 $\times$ 10$^{-16}$ g cm$^{-3}$, or number density $n_{0}$ = 4.2 $\times$ 10$^{8}$ cm$^{-3}$ (assuming a hydrogen atmosphere).  

As the model progresses, we keep track of the velocity of each ring, $v_{ij}$, the mass of red giant wind swept up by each ring $m_{ij}$ (note we will use lower case $m$ to refer to the swept up material, and upper case $M$ to refer to the ejecta), and the volume that the mass swept up by each ring occupies,  $V_{ij}$.  Both $m_{ij}$ and $V_{ij}$ are initially set to zero.  In a given time step, we advance each ring a distance $\Delta r_{ij} = v_{ij}t_{step}$, where $v_{ij}$ is the velocity of the $i^{th}$ ring and $t_{step}$ is the length of one time step, set to 1 hour.  Therefore, the $i^{th}$ ring will sweep up a mass  
\begin{equation}
\Delta m_{ij} = \rho_{ij} \int^{R_{ij}+\Delta r_{ij}}_{R_{ij}} \int^{\theta_{i+1}}_{\theta_{i}} \int^{2\pi}_{0} r^{2} sin(\theta) dr d\theta d\phi,                     
\end{equation} during the $j^{th}$ time step, assuming that the wind density encountered by each ring is approximately constant over the distance $\Delta r_{ij}$.  The total mass swept up by the $i^{th}$ ring since the start of the outburst, $m_{ij}$, is therefore given by 

\begin{equation}
m_{ij} = m_{i,j-1} + \Delta m_{ij}.
\end{equation}  As the rings of ejecta travels outwards, they each sweep up and shock the ambient medium they encounter.  In doing so, they transfer some of their kinetic energy to the swept-up material, and so the rings decelerate over time.  If we make the simplifying assumption that no mixing of the ejecta and the swept up material occurs, then $M_{i}$, the mass of a given ring of ejecta, remains constant with time.  At the start of the simulation, we assume that the evolution of the ejecta-environment interaction is {\it energy conserving}, such that 

\begin{equation}
\left (M_{i}+m_{ij} \right )v_{ij}^{2} = \left (M_{i}+m_{i,j-1} \right )v_{i,j-1}^{2}.
\end{equation}  At each time step, the velocity of the $i^{th}$ ring is recalculated using the above relationship.  The post-shock temperature of the swept-up gas, $T_{ij}$, is related to $v_{ij}$ by 

\begin{eqnarray}
T_{ij} & =  & \frac{3}{16} \frac{\mu m_{H}}{k} v_{ij}^{2} \nonumber \\
                   & =  & 1.4 \times 10^{7} \left (\frac{v_{ij}}{1000\ {\rm km\ s^{-1}}} \right)^2 K.
\end{eqnarray} At the earliest stages of the outburst, the swept-up mass is much lower than the ejected mass, and therefore little deceleration of the ejecta occurs.  This stage can be considered the free expansion phase of the outburst evolution.  Once the swept-up mass equals the ejected mass, the deceleration becomes much more pronounced, and at this point the outburst is in the Sedov phase of evolution.  We assume that the energy conserving phase lasts until the temperature of the post shock swept up gas drops to 10$^{6}$ K, at which point the swept-up material can cool efficiently via radiation.  We refer to this stage of evolution as the radiative phase, and at this point the ejecta--wind interaction is no longer energy conserving, but {\it momentum conserving}, such that

\begin{equation}
\left (M_{i}+m_{ij} \right )v_{ij} = \left (M_{i}+m_{i,j-1} \right )v_{i,j-1}.
\end{equation}
Since gas with temperatures below 10$^{6}$ K emits very little X-ray flux, we assume that swept-up material in the radiative phase does not contribute to the X-ray emission.  

In each time step, we calculate the emission measure of the material swept up by each ring using Equation 1.  This material occupies the region between the discontinuity surface $R_{ij}$, and the forward shock front, which we assume is at a distance of 1.2$R_{ij}$ from the white dwarf \citep[see e.g.][]{Chevalier82}.  The factor of 1.2 is appropriate for ejecta expanding into an $\rho \propto r^{-2}$ profile medium.  The associated volume of this region, which we denote $V_{ij}$, is given by 

\begin{equation}
V_{ij} = \int^{1.2R_{ij}}_{R_{ij}} \int^{\theta_{i}}_{\theta_{i+1}} \int^{2\pi}_{0} r^{2} sin(\theta)\ dr\ d\theta\ d\phi.
\end{equation}

Assuming a constant density across the forward shock region, the number density of the swept up material is given by 

\begin{equation}
n_{ij} = m_{ij}/\mu m_{H} V_{ij}, 
\end{equation} recalling that $m_{ij}$ is the {\it total} mass swept up by the $i^{th}$ ring since the onset of the outburst. This can be inserted into Equation 1 to obtain the emission measure:

\begin{equation}
EM_{ij} = n_{ij}^{2} V_{ij}.
\end{equation}  Finally, we can combine this emission measure and the shock temperature in Equation 9 to obtain an estimate of the X-ray luminosity of each swept up ring using the following approximation:
\begin{eqnarray}
L_{ij} & =  & 1.4 \times 10^{-27}\ T_{ij}^{0.5} n_{e} n_{i} Z^{2} \bar{g}_{B}\ V_{ij} \nonumber \\
           & \sim  & 1.4 \times 10^{-27}\ T_{ij}^{0.5} EM_{ij}.
\end{eqnarray} where we have assumed a completely ionized hydrogen plasma and $\bar{g}_{B}$ is the Gaunt factor, $\sim$1 \citep[see Equation 5.15b, ][]{Rybicki79}.

The equations laid out thus far calculate the shock properties for each individual ring of ejecta.  We of course have no observational capability to spatially resolve the ejecta as they interact with the giant wind.  Therefore, we must estimate global parameters that can be compared with the data by combining the parameters from all rings at each time step.  The total emission measure $\mathcal{EM}_{j} $ and luminosity $\mathcal{L}_{j}$ are given by a simple summation over all rings:

\begin{equation}
\mathcal{EM}_{j} = \sum^{180}_{i=1} EM_{ij},
\end{equation}

\begin{equation}
\mathcal{L}_{j} = \sum^{180}_{i=1} L_{ij}.
\end{equation}

The observed temperature, $\overline{\mathcal{T}}_{j}$, is an average over all rings weighted by their emission measure:

\begin{equation}
\overline{\mathcal{T}}_{j} = \frac{\sum^{180}_{i=1} EM_{ij} T_{ij}}{\sum^{180}_{i=1} EM_{ij}}.
\end{equation} We track these global parameters over the course of the model run, and plot these quantities as a function of time.  Two examples of these output plots are shown is Figure 8.  We also produce a visualization of the spatial evolution of the ejecta by plotting $R_{ij}$ for all $\theta_{i}$ at intervals of 100 hours.  We color code each surface depending on what phase (free expansion, Sedov or radiative) the ejecta are in.  Examples of this are given in Figure 9.  Note that for the runs shown in Fig. 9, the expansion of the ejecta never becomes radiative.

\begin{figure*}
\begin{center}
\includegraphics[width=2.3in,angle=90]{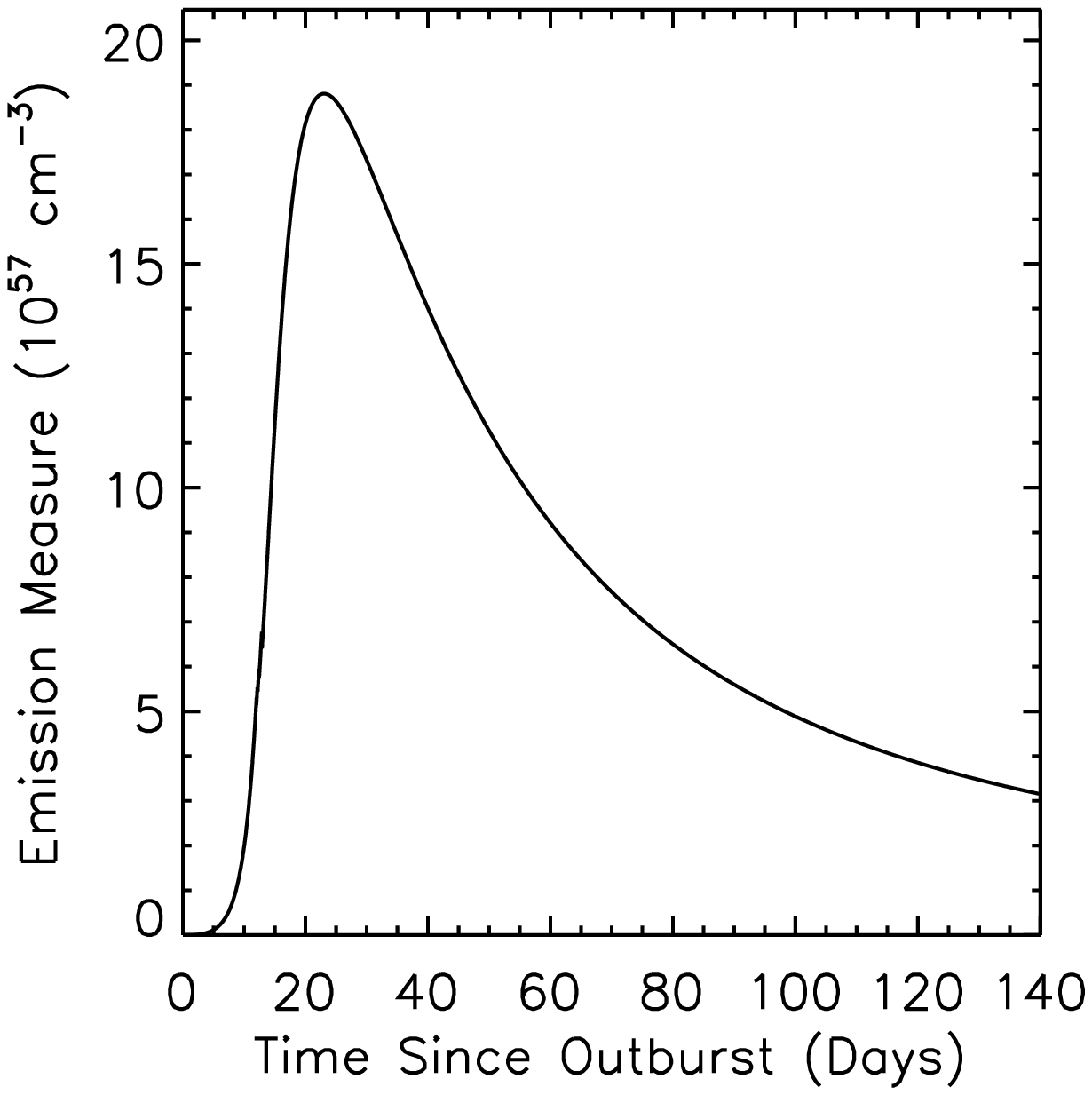}
\includegraphics[width=2.3in,angle=90]{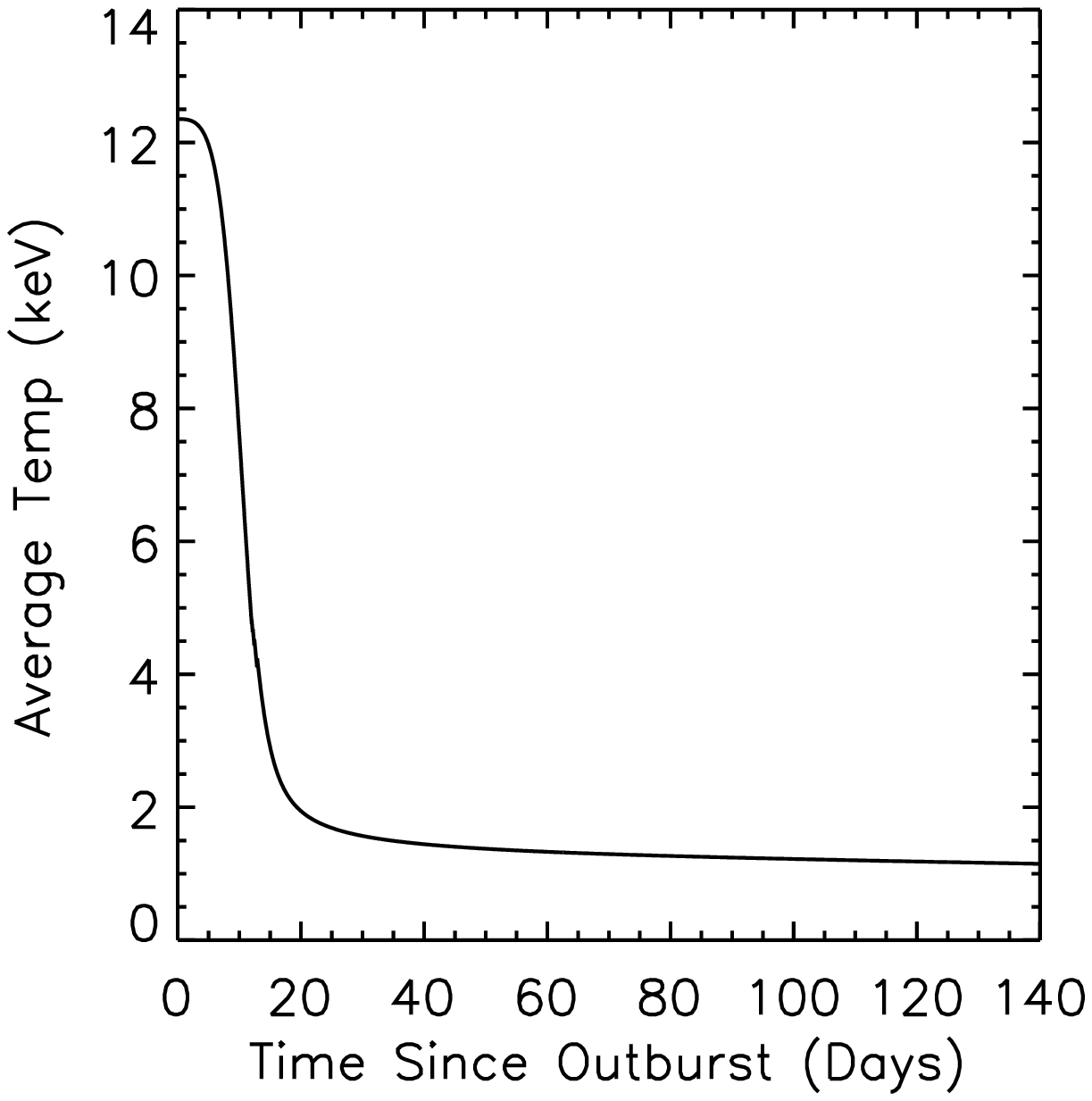} 
\includegraphics[width=2.3in,angle=90]{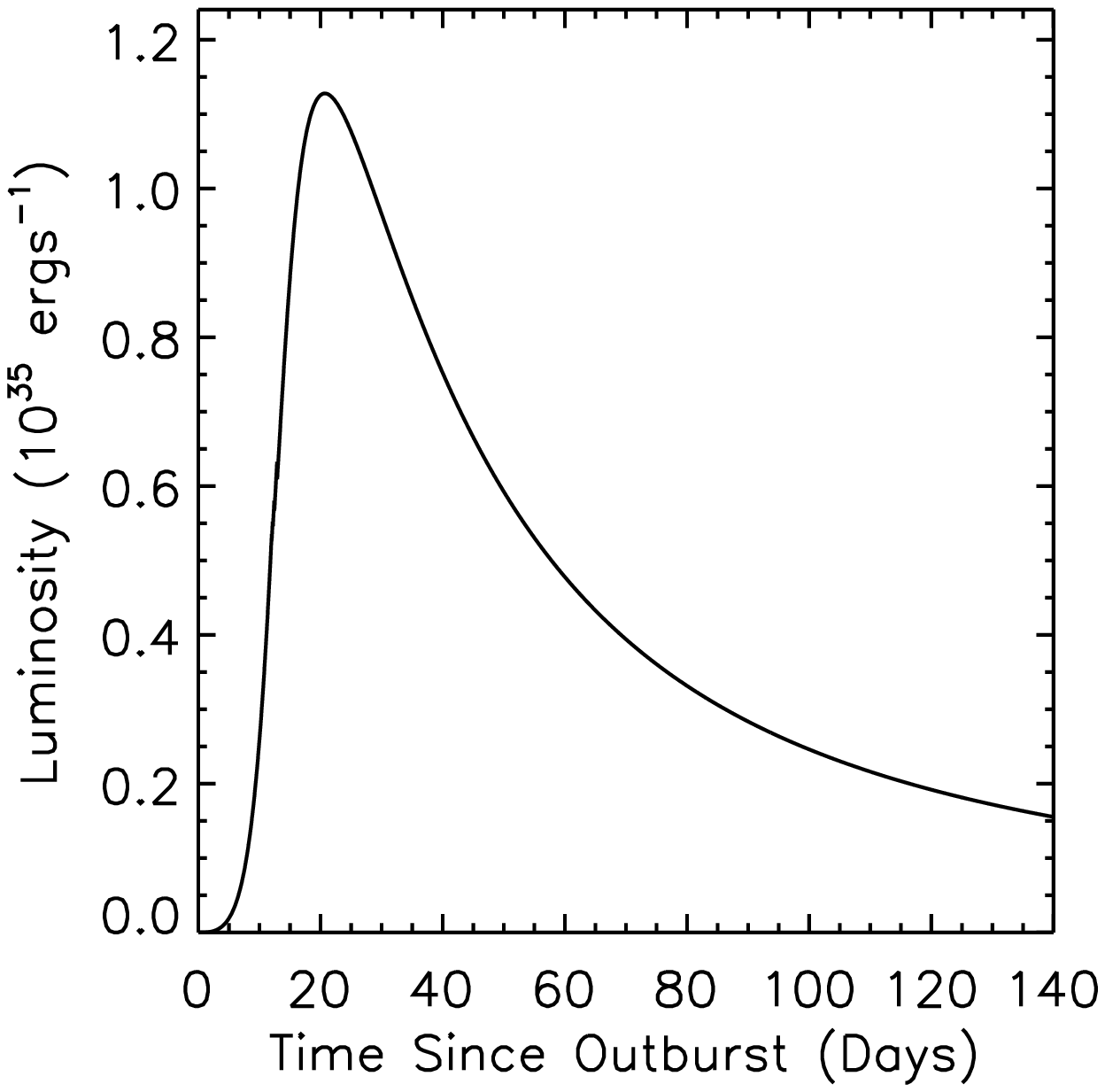} \\
\includegraphics[width=2.3in,angle=90]{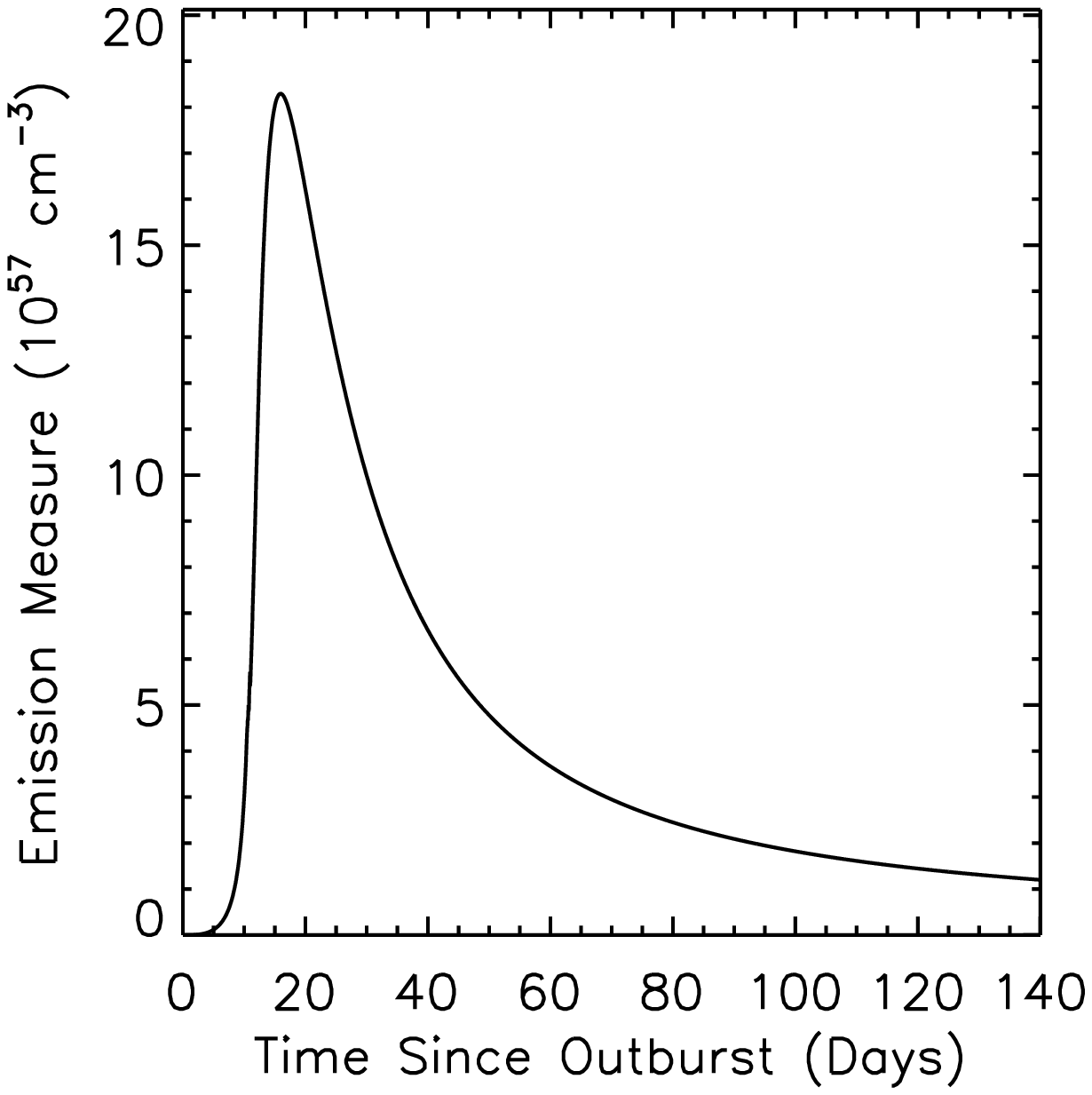}
\includegraphics[width=2.3in,angle=90]{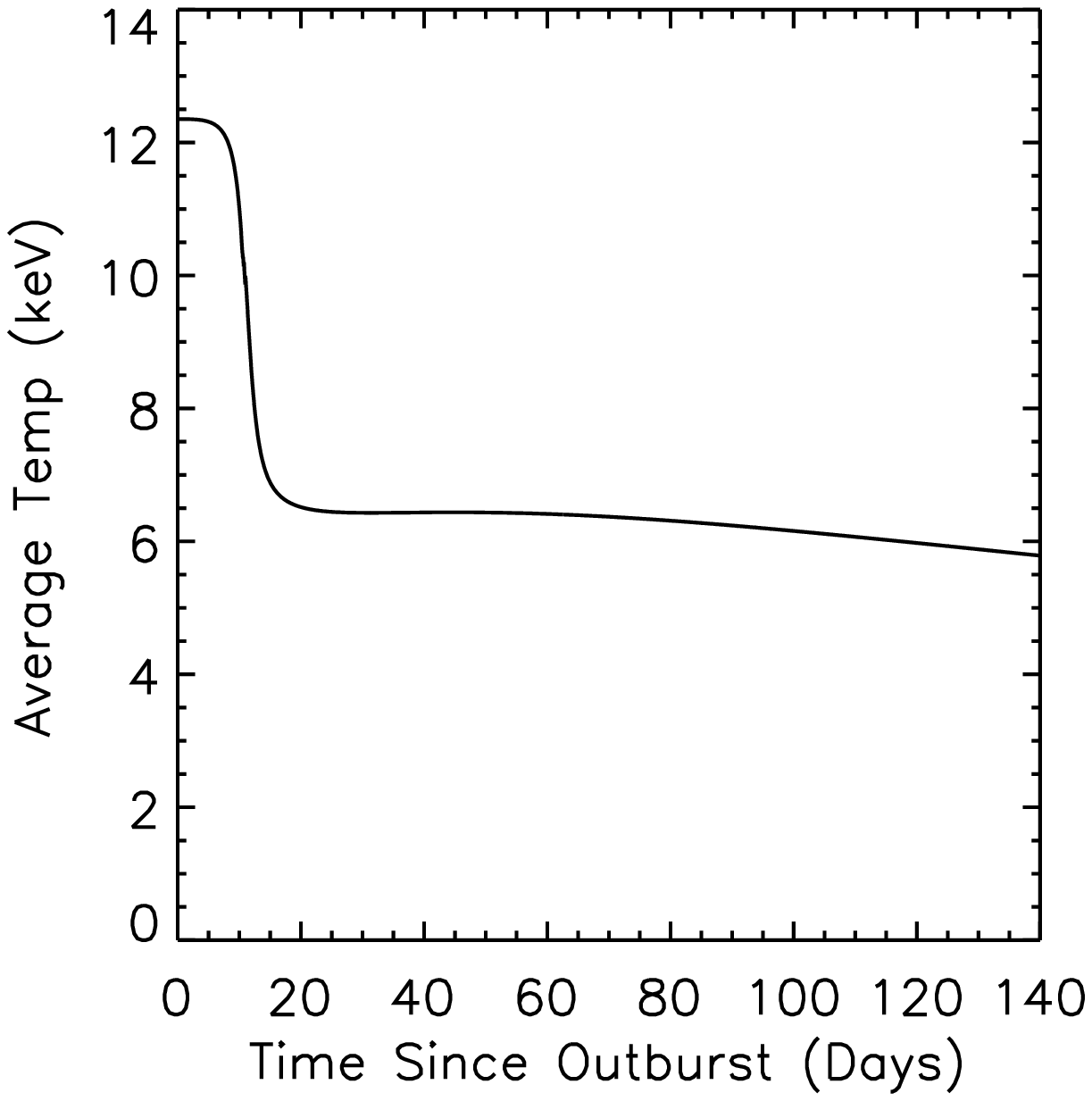}
\includegraphics[width=2.3in,angle=90]{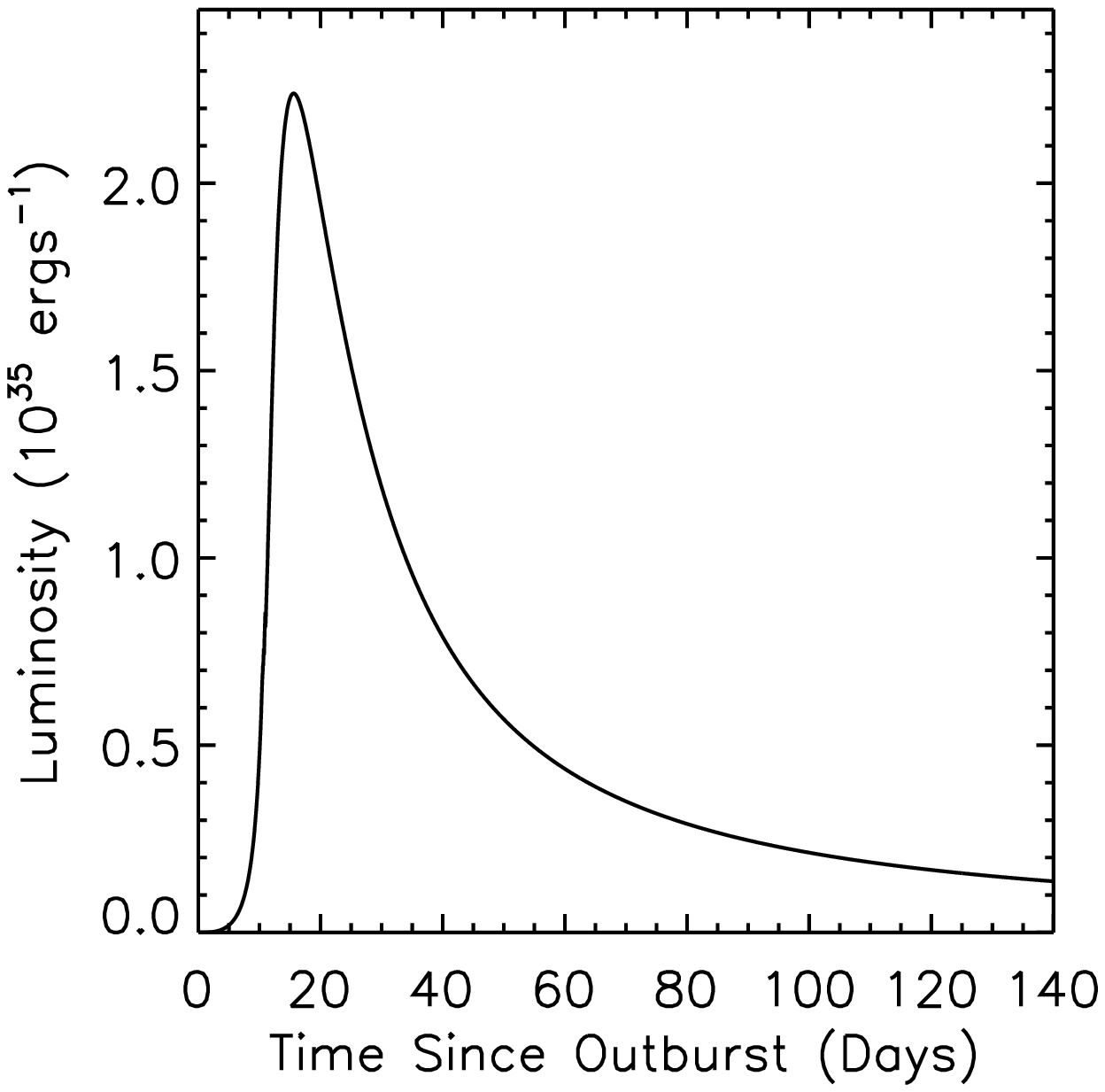}
\caption{Example output of the simple model described in Section 6, showing the evolution of the forward shock emission measure ({\it left}), emission measure weighted temperature ({\it middle}) and luminosity ({\it right}).  These two model runs assumed a binary separation of 25 AU and a mass loss rate from the giant of 10$^{-7}$ M$_{\odot}$ yr$^{-1}$. In the top row, the nova ejected mass is 5 $\times$ 10$^{-7}$ M$_{\odot}$, and in the bottom it is 5 $\times$ 10$^{-6}$ M$_{\odot}$. }
\label{fig6}
\end{center}
\end{figure*}

\subsection{Modeling results}
We present the results of our fiducial model, assuming an ejected mass of either 5 $\times$ $10^{-7}$ M$_{\odot}$ or 5 $\times$ $10^{-6}$ M$_{\odot}$ (typical of ejecta masses in fast novae, see, e.g. Yaron et al. 2005), in Figures 8 and 9.  We also explored a small parameter space of nova ejecta mass, binary separation and wind mass loss rate, since none of these parameters are tightly constrained.  Independent of the assumed binary separation, ejecta mass and red giant mass loss rate, we see some gross features in our model results which are common to all parameter sets.

The emission measure and luminosity are initially very low.  As the ejecta emitted close to $\theta$ = 0$^{\circ}$ approach the surface of the giant companion, these parameters increase rapidly and dramatically to a peak value several orders of magnitude higher than at the start of the simulation.  Then, once all ejecta are moving outward from the central binary, the emission measure and luminosity begin to drop, although at a rate which is decreasing with time.

\begin{figure*}
\begin{center}
\includegraphics[width=3in,angle=90]{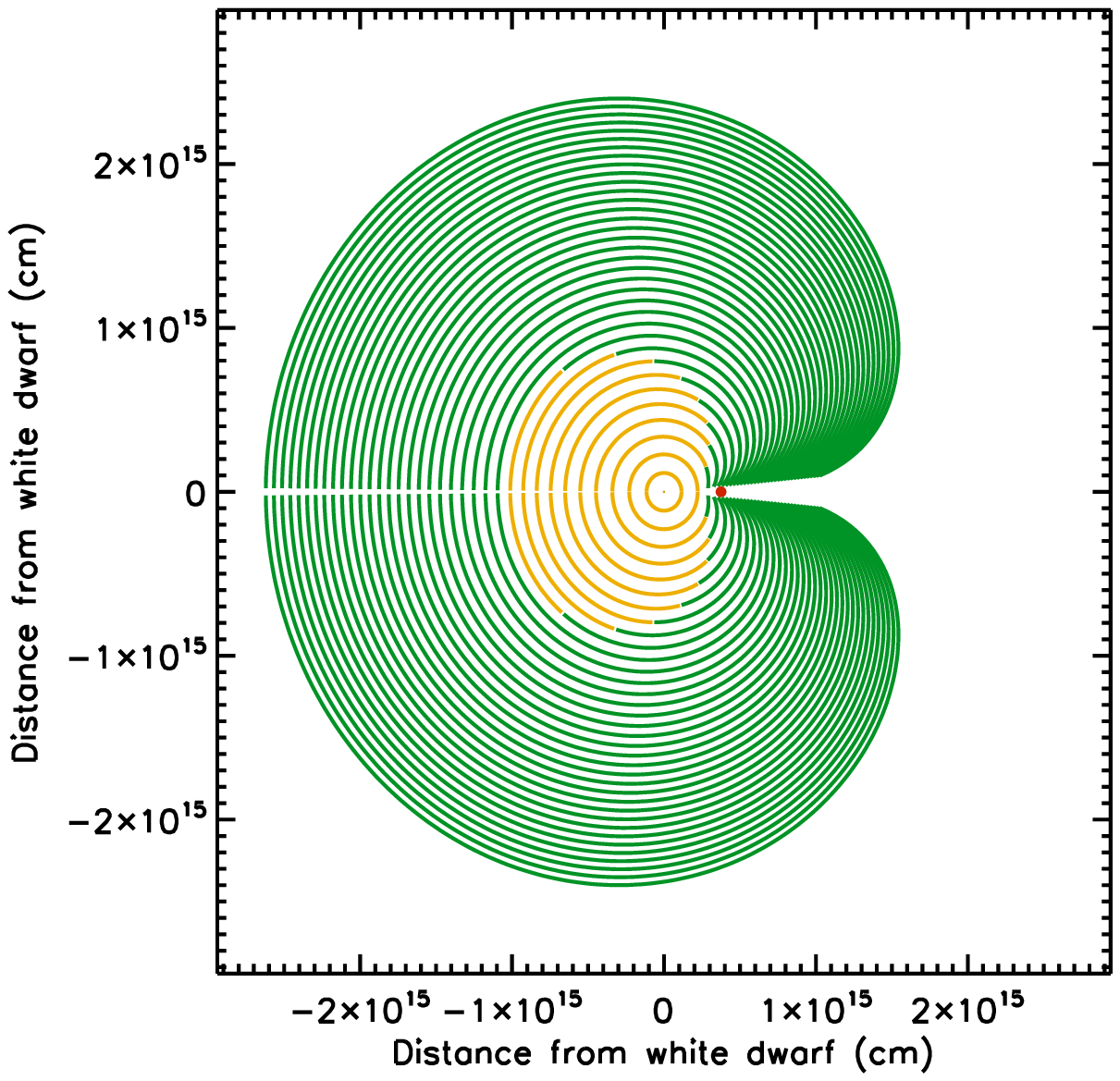} 
\includegraphics[width=3in,angle=90]{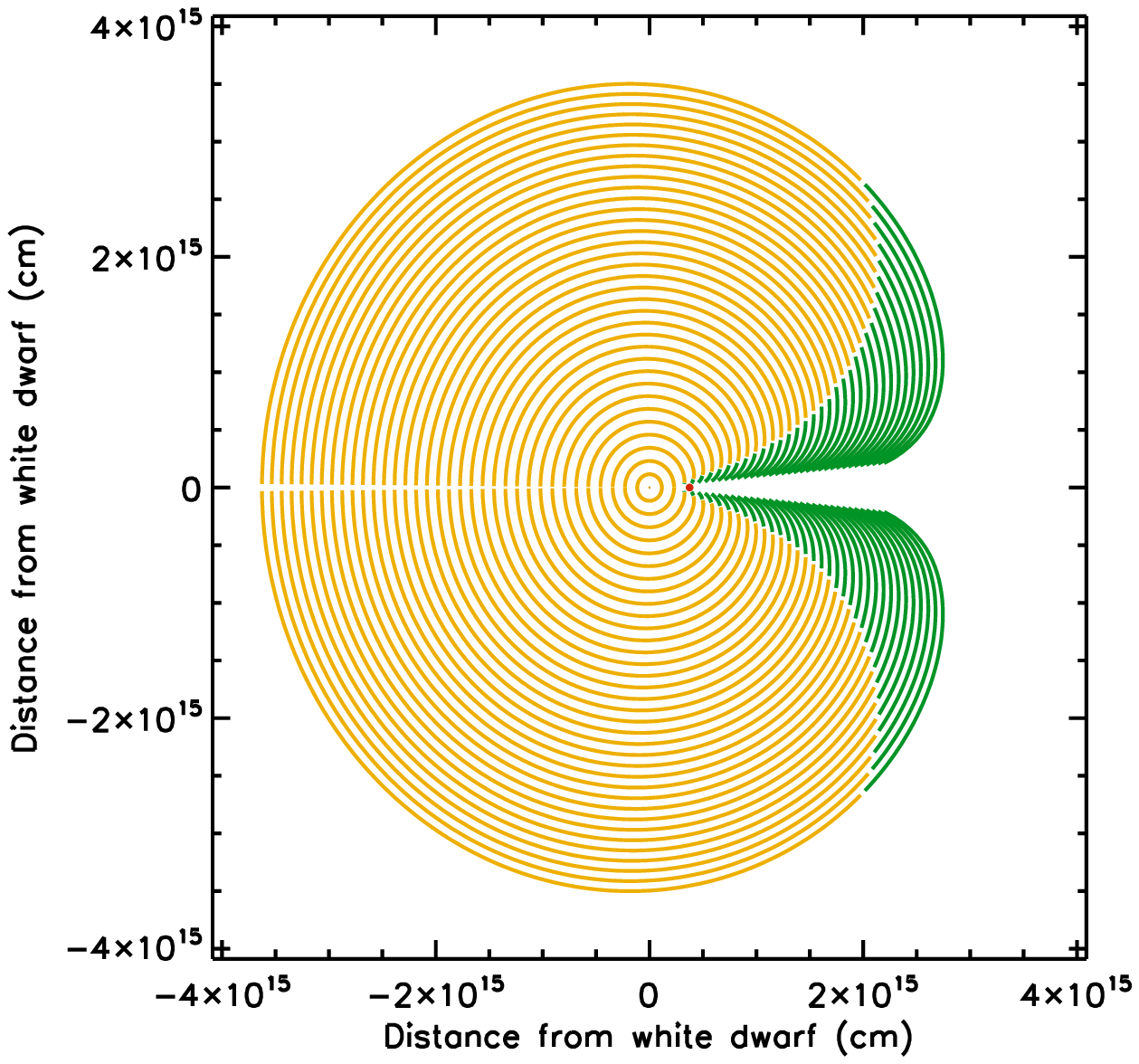}
\caption{Example of the evolution of the size of the nova ejecta over the first $\sim$200 days of the outburst, for an ejected mass of 5 $\times$ 10$^{-7}$ M$_{\odot}$ ({\it left}) and 5 $\times$ 10$^{-6}$ M$_{\odot}$ ({\it right}).  Each line corresponds to an increase of 100 hours since the onset of the outburst.  The red circle represents the giant, and its radius is to scale.  Gold lines correspond to regions which are in free expansion, green lines to regions which are in the Sedov (energy conserving) phase.  The same binary separation and wind mass loss rate are assumed as in Figure 8.}
\label{fig7}
\end{center}
\end{figure*}

The temperature on the other hand starts off at a high value set by the initial ejecta velocity.  While all the ejecta are in the free expansion phase, this temperature remains approximately constant.  However, the weighted mean value begins to drop rapidly around the same time as the luminosity increases, since the majority of the emission comes from gas very close to the red giant, which has slowed the ejecta down significantly due to its high density.  The temperature then stays approximately constant for the remainder of the simulation.  These features can be compared with the evolution of the luminosity and temperature observed in the \textit{Swift} data presented in Figures 5.  For the models presented in Fig 8, we find close agreement in terms of the shape of the lightcurve and the observed temperatures.  The observed temperature drop is steeper in the model than the observe in the data during the first 10--20 days.  The most obvious discrepancy between the model and the data is found in the emission measure and luminosity---the model predicts values that are an order of magnitude larger than we see in the data.  This, and the incorrect rate of temperature change, are probably due to some oversimplifications in our model assumptions that we will discuss in the next section.

The choice of input binary separation, ejected mass and mass loss rate do not alter the overall shape of our curves, but do result in significant changes in the onset of the X-ray rise, the peak emission measures and luminosities reached, the post shock temperatures, and the rate of decay from maximum.   We find that for binary separations less than 20 AU, the peak X-ray flux was reached at much earlier times than observed in the data, in agreement with the binary separations implied by the presence of HBB in the Mira component (see discussion in Section 1.2).   Figure 8 demonstrates the effect of changing the ejected mass for a single choice of binary separation and red giant mass loss rate.  Increasing the ejecta mass from 5 $\times$ 10$^{-7}$ to 5 $\times$ 10$^{-6}$ M$_{\odot}$ results in the time of peak X-ray brightness occurring $\sim$1 week earlier.  This is because the higher initial kinetic energy of the ejecta results in less deceleration by the ambient wind, and thus a faster crossing of the binary system.  Note that the peak emission measures are similar, since this is primarily set by the density structure of the red giant wind.

There are significant differences in the plasma temperature at late times, with the value trending towards $\sim$6 keV for the higher ejecta mass, rather than 2 keV in the 10$^{-7}$ M$_{\odot}$ case.  Again, this is because the smaller deceleration of the ejecta by the wind results in larger ejecta velocities and hence higher post-shock temperatures.  The higher temperature therefore results in a slightly higher X-ray luminosity.  A faster decay from maximum is also found in the model with higher ejecta mass, presumably because the ejecta are moving more quickly as they pass the red giant and expand out into the larger environment. 

Figure 9 shows the evolution of the size of the ejecta over time for the two ejected mass cases.  Gold lines indicate that the ejecta are still effectively in free expansion, i.e. they have not yet swept up their own mass.  Green lines indicate ejecta that are in the Sedov phase of evolution.  The effect of the higher mass is obvious - the ejecta remain in the free expansion phase for longer, since it takes more time to sweep up the ejected mass.  The ejecta are also larger in volume, since they have a higher velocity for a longer period of time.  

\subsection{Limitations of the model}
Our simple model neglects some important physics.  First,  we only consider the evolution of the forward shock, and do not track in any way the reverse shock being driven back through the ejecta.  Second, we consider only a simple strong shock with no mixing of the ejecta and the swept up gas.  In reality, important instabilities form which lead to mixing across the discontinuity surface, and regions of high density enhancements which can result in rapid cooling of portions of the swept up material.  Furthermore, we assume that the gas in the swept up shell is  thoroughly mixed on a time scale much smaller than one time step in our model, so that its density is uniform and always given by Equation 12.  In reality this is not the case, and the swept up shell likely has some density profile in the radial direction away from the white dwarf.  Such a profile will have implications for the value of the emission measure, and for how cooling proceeds within the shell.  In fact, it  is likely that the very earliest material swept up by rings traveling away from the giant cools within a time period shorter than our model run.  Therefore, assigning this material the shock temperature for later material is incorrect, and probably biases us to larger values of the emission measure and luminosity. 

Thirdly, we assume that the forward shock is located at distance 1.2$R_{ij}$ from the white dwarf.  This value is taken from \citet{Chevalier82}, and assumes that the ejecta are centered at the origin of an $r^{-2}$ density profile.  This assumption clearly breaks down for material moving towards the red giant at early times, and in fact for this material the forward shock is likely located closer to the discontinuity surface.  In these directions, our emission measures are probably underestimated as the swept up material in fact occupies a smaller volume.  

Finally, we neglect the effects of particle acceleration on the temperature of the post-shock gas.  The detection of V407 Cyg with Fermi during the first few weeks is clear evidence that some of the shock energy was imparted into relativistic particles.  \citet{Tatischeff07} discuss the importance of this effect on the early evolution of RS Oph, and suggest that the diffusive shock acceleration of particles was responsible for the observed rate of deceleration of the ejecta, which was higher than predicted by standard shock wave models.  The primary piece of observational evidence for this effect are X-ray temperatures lower than expected from the observed line velocities using Equation 9.  Thus, our model may over-predict the post-shock  temperature at early times.  If we take the presence of GeV emission to indicate that diffusive shock acceleration is an important source of cooling, then we would expect to see lower X-ray temperatures during the first two weeks of the outburst.  Unfortunately, this is the epoch where the X-ray emission is very faint, and we cannot determine from the existing data if this was an important source of cooling.  

Clearly, the assumptions we have made have implications for the parameter values (emission measure, temperature etc) implied by our simple model.  However, we note that regardless of the exact details of the physics, most of the of X-ray emission in our model originates in the material swept up by the ejecta traveling towards the red giant.   
 
\section{Comparison with RS Oph}
To date, RS Oph is the only other nova with a giant companion which has been well observed in X-rays \citep[see, e.g.][]{Bode06,Sokoloski06}.  Here, we contrast and compare their X-ray evolution. During the 2006 outburst of RS Oph, the Burst Alert Telescope (BAT) onboard the {\it Swift} satellite obtained a statistically significant detection of the system in the 14--25 keV range in the first five days.  Despite the detection of high energy gamma rays with {\it Fermi}, no similar BAT detection was made during the first few days of the V407 Cyg outburst.  

For both systems, the first {\it Swift} XRT observations were obtained on day 3 of the outburst, with subsequent observations occurring every few days at a similar rate.  In the case of RS Oph, the first pointing was extremely bright, with a 0.3--10 keV XRT count rate of 14.2 $\pm$ 0.2 counts s$^{-1}$.  Even allowing for a greater distance to V407 Cyg, this is still several orders of magnitude brighter than in V407 Cyg (with 0.011 $\pm$ 0.004 counts s$^{-1}$ on day 3).  Furthermore, RS Oph reached its maximum 0.3-10 keV count rate of 31.5 $\pm$ 0.2 around day 5, and then faded in all subsequent observations.  Given that the similar initial ejecta velocities (4000 km s$^{-1}$ for RS Oph vs 3200 km s$^{-1}$ for V407 Cyg) imply shock temperatures that differ by only a factor of 1.5, and that the intrinsic column densities are also similar, then the large difference in X-ray count rate must be due to the very different density of the material in the immediate vicinity of the white dwarf.  

The most striking difference between the lightcurves of RS Oph and V407 Cyg is the emergence of an extremely bright supersoft source in RS Oph \citep{Nelson08,Ness09}, believed to originate in the still hydrogen burning shell on the surface of the white dwarf.  In RS Oph, this component emerged about 25 days into the outburst, and then turned off quite suddenly around 50 days later.  While we do detect a blackbody like component with a similar temperature in V407 Cyg, its observed count rate is much lower.  Fits to the {\it Swift} spectra presented in \citet{Bode06} show that the intrinsic absorption had dropped to a few 10$^{21}$ cm$^{-2}$ by day 10 of the 2006 outburst in RS Oph.  In stark contrast, the intrinsic absorption in our model fits to the V407 Cyg {\it Suzaku} spectrum on day 30 is $\sim$1.5 $\times$ 10$^{22}$ cm$^{-2}$, almost an order of magnitude higher.  Therefore, the lack of a bright supersoft phase in the outburst evolution of V407 Cyg is due to the much higher intrinsic absorption and soft X-ray attenuation than in RS Oph.

\section{Discussion}
\label{discussion}

\subsection{The role of environmental asymmetry in the X-ray evolution of the outburst}
The observed brightening of the lightcurve around day 30 can be explained by the approach of the ejecta towards the red giant photosphere.  Even though only a small fraction of the ejecta intersects the red giant, the large increase in density towards the base of the red giant wind results in a dramatic increase in the emission measure of the shell, and therefore of the X-ray luminosity.  The rest of the shell emits at a lower luminosity and higher temperature, as deceleration of the ejecta is less efficient in these directions.  This is reflected in our best fit {\it Suzaku} model, where we find that $\sim$80\% of the 2--10 keV flux originates in a 2.5 keV plasma that is already in CIE, and the remaining 20\% of the flux is due to the hotter NEI plasma, with a temperature of $\sim$3.9 keV.  We examined the output of our model, and found that by dividing the emitting shell into two regions contributing 80\% and 20\% of the flux, the emission measure averaged temperatures of these two regions were similar to those found in our spectral fits.  Furthermore, the region emitting 80\% of the flux was within 20 degrees  of the line of sight between the two stars, i.e. a small fraction of the total swept up shell.

The presence of both NEI and CEI plasmas is firm observational evidence of the direction dependency of the density profile encountered by the nova ejecta.  According to \citet{Smith10}, most species will be in collisional equilibrium with an ionization age $\tau$ = 10$^{12}$ cm$^{-3}$s.  In the vicinity of the red giant, the density is high and plasma comes into equilibrium extremely quickly.  In the opposite direction, the density is lower, and the time to come into equilibrium is longer.  Also, since the density in this direction drops as the ejecta move outwards, the equilibrium timescale will decrease with time.  The presence of NEI plasma over the course of our 40 ks observation, and a best fit ionization age of $\sim$2.2 $\times$ 10$^{10}$ cm$^{-3}$s, implies a maximum density of 5 $\times$ 10$^{5}$ cm$^{-3}$ for the plasma away from the red giant.  Assuming the same fiducial values for the red giant radius and mass loss rate as in Section 6, this density implies a distance from the red giant center of $\sim$10$^{15}$ cm, or 67 AU.  Assuming little deceleration in the direction away from the red giant, and an initial ejecta velocity of 3200 km s$^{-1}$, then the outermost parts of the ejecta should be about 55 AU away from the white dwarf surface.  Adding the binary separation of at least 20 AU, then ejecta should be $\sim$75 AU away form the red giant center.  Thus the distances are in approximate agreement, and it is reasonable that we see an NEI component at day 30.     

\subsection{A massive white dwarf in V407 Cyg}
Several features of the optical, UV and X-ray evolution provide evidence that the white dwarf in V407 Cyg is very massive.  The optical evolution was fast, typical of novae on more massive white dwarfs (like the recurrent novae).  Our model fits to the {\it Swift} and {\it Suzaku} spectra indicate blackbody temperatures in the range 40-60 eV, a range very similar to that observed in RS Oph \citep{Osborne11} which is known to host a massive white dwarf.   The optical and X-ray properties of V2491 Cyg suggest that it may also be a recurrent nova hosting a massive white dwarf \citep{Ness11,Darnley11}. The blackbody temperatures found for the {\it Swift} spectra of V2491 Cyg were somewhat higher, averaging $\sim$70 eV \citep{Page10}. Despite the uncertainties in the model fits, and the problems of using blackbodies to model novae in outburst, the presence of flux between 0.5 and 0.8 keV associated with a soft spectral component seem to favor an emitting source with characteristics similar to both RS Oph and V2491 Cyg.

Considering the X-ray and UV lightcurves in tandem, we note that the break observed in the optical and UV around day 50 coincides with the turning off of the soft component in the X-ray spectrum.  A similar behavior was observed in RS Oph by \citet{Hachisu07}, who observed a break in the optical lightcurve (in that case, the Stromgren y-band filter) around day 75.  At the same time, a rapid fading of the supersoft emission began.  The authors attribute this behavior to the cessation of nuclear burning on the white dwarf surface.  If we assume that the observed break and corresponding turn off of the soft component in V407 Cyg have the same origin, then the short turn off time ($\sim$50 days) points to a nova outburst on a very massive white dwarf.  More massive white dwarfs have higher surface gravities and therefore reach the critical conditions for a surface thermonuclear runaway with much smaller shell masses.  

Finally, the presence of Li in the spectrum of the AGB component in V407 Cyg is indicative HBB.  \citet{Mikolajewska10} points out the HBB only occurs in stars with initial masses in the range 4--8 M$_{\odot}$, implying that the Mira variable in V407 Cyg is an intermediate mass star.  Since the white dwarf is more evolved, its progenitor must have been more massive than the Mira companion.  Therefore, by evolutionary considerations alone, we would expect a massive white dwarf in this system.  The observed properties of the outburst appear to be in agreement with this expectation.

Other symbiotic stars hosting massive white dwarfs (e.g. T CrB, RT Cru and CH Cyg) have been detected at energies above 10 keV with hard X-ray instruments onboard {\it Swift, Integral} or {\it Suzaku} \citep[see, e.g.][]{Kennea09}. In general, their spectra are well described by cooling flow models with maximum temperatures between 30 and 70 keV.  This emission originates in shocked gas in the accretion disk boundary layer, where disk material must shed its excess energy before accreting onto the white dwarf surface.  The characteristic temperature of this emission increases with white dwarf mass.  If similar high temperature emission was detected in future X-ray observations of the system in quiescence, this would be further proof of a high mass system (although we note the counter-example of RS Oph, where despite the known high mass of the white dwarf, the boundary layer emission in quiescence has a temperature of only 6 keV.  For further details, see the discussion in \citet{Nelson11}).

\subsection{Why the X-ray rise and $\gamma$-ray decline are temporally coincident}
\citet{Abdo10} first noted the striking temporal coincidence between the cessation of the $\gamma$-ray emission observed with \textit{Fermi}, and the rapid brightening of the X-ray flux observed with \textit{Swift}.  Our detailed X-ray analysis and simple modeling of the outburst have identified the sweeping up of material {\it in the immediate vicinity of the red giant} as the source of rapid X-ray rise.  This is the opposite of the suggestion in Abdo et al. that the X-rays rise due to ``the increasing volume of shocked gas in the nova shell expanding in the direction away form the red giant".  Knowing that a fraction of the ejecta have reached the red giant at the time of the gamma ray decline allows us to propose a scenario for the gamma ray turn off.

The GeV gamma-ray emission observed with {\it Fermi} is produced by the interaction of a population of particles accelerated to high energies in the shock front with a second group of particles \citep{Abdo10}.  In the leptonic scenario, accelerated electrons up-scatter infrared photons to GeV energies through the inverse Compton process.  In the hadronic scenario, accelerated protons interact with slow protons in the swept up medium, producing neutral pions which subsequently decay and emit GeV photons.  A detectable signal therefore requires a combination of  1) a shock speed that can accelerate particles up to energies high enough to produce GeV photons through these secondary interactions and 2) a number of accelerated particles and subsequent interactions sufficient to produce a high gamma ray flux.  

In directions away from the giant, the density encountered by the ejecta decreases.  Therefore, the number of accelerated particles will drop with time, {\it even if the shock velocity remains high and can still accelerate particles to high energies}.  Furthermore, the density of target particles is also dropping off with time for both the leptonic and hadronic production scenarios, since the radiation field and the Mira wind both have $1/r^{2}$ density profiles.  In combination, these two effects result in a rapidly decreasing gamma ray flux as the ejecta travel out from the white dwarf.

In directions towards the red giant, the density of the wind and the radiation field is increasing, and so the number of accelerated particles and targets will grow with time.  While the shock velocity remains high (i.e. while little mass is swept up), the gamma ray flux remains high.  However, as the ejecta approach the red giant, they experience a rapidly increasing deceleration as more and more red giant wind is swept up.  This drop in shock velocity reduces the maximum energy that the particles can be accelerated to.  This will cause the GeV flux to decrease rapidly, despite the fact that both the number of accelerated particles {\it and} targets is increasing.  As the shock stalls, there is simply insufficient energy available to power the production of GeV photons.  This makes it likely that most of the gamma rays were produced in the material swept up by ejecta moving towards the red giant.  We note that while we have presented a plausible explanation for why the coincidence in the gamma ray decline and the the X-ray rise based on our improved understanding of the X-ray evolution, we cannot use this new insight to favor a leptonic production mechanism over a hadronic one for the gamma rays.  This is because the radiation field and the Mira wind have the same density profile.  

Given the requirements outlined here for a significant gamma ray flux, we expect that only nova outbursts in wide binaries, and with giant companions, will ever produce gamma ray emission of the type observed in V407 Cyg.  The high density environment provided by the giant wind is crucial for supplying the required large number of gamma-ray producing interactions.  The wide binary separation is necessary to allow time for the system to be detected - too small a separation and the ejecta will decelerate very quickly, making detection with an instrument like {\it Fermi} unlikely.   In light of the rarity of symbiotic stars, gamma rays from novae will be extremely rare, making the detection of V407 Cyg with  {\it Fermi} a truly extraordinary event. 

\section{Conclusions}
\label{conclusions}
We have presented X-ray observations obtained with \textit{Swift} and \textit{Suzaku} during the 2010 outburst of the symbiotic Mira V407 Cyg.  The X-ray emission is initially faint, and then brightens rapidly approximately two weeks after the start of the outburst.  A detailed spectroscopic analysis of the X-ray emission at peak brightness reveals the existence of two distinct emitting components.  The first dominates between 2 and 10 keV, and originates in the forward shock being driven into the Mira wind by the ejecta.  This emission is itself quite complex.  We find evidence for two different plasmas, one in collisional ionization equilibrium with $kT$ = 2.3 keV, contributing $\sim$80 \% of the observed flux, and one non-equilibrium ionization plasma with ionization age $\tau$ = 2.2 $\times$ 10$^{10}$ cm$^{-3}$ s and $kT$ = 4 keV, contributing the remaining 20\% of the flux.  The presence of an NEI plasma component is direct observational evidence of the asymmetry in the environment encountered by the nova ejecta.

The second component peaks at energies below 1 keV, and we identify it as supersoft X-ray emission from the white dwarf seen through the dense Mira wind.  This emission appears to turn off around day 50, coincident with a change in the slope of the optical and UV lightcurves, and therefore most likely indicates the cessation of nuclear burning on the white dwarf.  Such fast turn off times are indicative of nova outbursts on massive white dwarfs, including the recurrent novae.  Therefore, V407 Cyg joins the set of symbiotics hosting massive white dwarfs. Future observations of a high-temperature X-ray spectrum in quiescence will provide an important test of this result.

We have developed a simple model of the ejecta-environment interaction, and have used it to show that the rapid rise and subsequent decay in the X-ray light curve is due to the ejecta approaching and passing the red giant.  As a result, the most likely explanation for the coincidental gamma-ray decline/X-ray rise is the stalling of the shock front as the ejecta approach the red giant photosphere.  We expect that GeV emission will only be detectable in nova events occurring in wide binaries with giant companions, and will therefore be extremely rare.  

\acknowledgements
We thank the {\it Swift} and {\it Suzaku} projects for the generous allocation of target-of-opportunity time that made this work possible.  We also thank Martin Laming and Cara Rakowski at the Naval Research Laboratory for interesting discussions about shocks.

\bibliography{v407cyg.bib}

\end{document}